%% file: main.tex
\documentclass[twocolumn]{aastex631}
\usepackage{enumitem}
\usepackage{amsmath}
\usepackage[mathlines]{lineno}

\shortauthors{Subrayan et al.}

\begin{document}

\title{Scary Barbie: An Extremely Energetic, Long-Duration 
Tidal Disruption Event Candidate \\ Without a Detected Host Galaxy at z = 0.995}

\author[0000-0001-8073-8731]{Bhagya M.\ Subrayan}
\affiliation{Purdue University, Department of Physics and Astronomy, 525 Northwestern Ave, West Lafayette, IN 47907, USA}

\author[0000-0002-0763-3885]{Dan Milisavljevic}
\affiliation{Purdue University, Department of Physics and Astronomy, 525 Northwestern Ave, West Lafayette, IN 47907, USA}
\affiliation{Integrative Data Science Initiative, Purdue University, West Lafayette, IN 47907, USA}

\author[0000-0002-7706-5668]{Ryan Chornock}
\affiliation{Department of Astronomy, University of California, Berkeley, CA 94720-3411, USA}

\author[0000-0003-4768-7586]{Raffaella Margutti}
\affiliation{Department of Astronomy, University of California, Berkeley, CA 94720-3411, USA}

\author[0000-0002-8297-2473]{Kate D.\ Alexander}
\affiliation{Department of Astronomy and Steward Observatory, University of Arizona, 933 North Cherry Avenue, Tucson, AZ 85721-0065, USA}

\author[0000-0002-9176-7252]{Vandana Ramakrishnan}
\affiliation{Purdue University, Department of Physics and Astronomy, 525 Northwestern Ave, West Lafayette, IN 47907, USA}

\author[0000-0001-7626-9629]{Paul C.\ Duffell}
 \affiliation{Purdue University, Department of Physics and Astronomy, 525 Northwestern Ave, West Lafayette, IN 47907, USA}

\author[0000-0003-0913-4120]{Danielle A.\ Dickinson}
\affiliation{Purdue University, Department of Physics and Astronomy, 525 Northwestern Ave, West Lafayette, IN 47907, USA}

\author[0000-0003-3004-9596]{Kyoung-Soo Lee}
\affiliation{Purdue University, Department of Physics and Astronomy, 525 Northwestern Ave, West Lafayette, IN 47907, USA}

\author[0000-0003-1503-2446]{Dimitrios Giannios}
\affiliation{Purdue University, Department of Physics and Astronomy, 525 Northwestern Ave, West Lafayette, IN 47907, USA}

\author[0000-0001-9314-0683]{Geoffery Lentner}
\affiliation{Purdue University, Department of Physics and Astronomy, 525 Northwestern Ave, West Lafayette, IN 47907, USA}

\author{Mark Linvill}
\affiliation{Purdue University, Department of Physics and Astronomy, 525 Northwestern Ave, West Lafayette, IN 47907, USA}

\author[0000-0001-6922-8319]{Braden Garretson}
\affiliation{Purdue University, Department of Physics and Astronomy, 525 Northwestern Ave, West Lafayette, IN 47907, USA}

\author[0000-0002-3168-0139]{Matthew J.\ Graham}
\affiliation{Division of Physics, Mathematics, and Astronomy, California Institute of Technology, Pasadena, CA 91125, USA}

\author[0000-0003-2686-9241]{Daniel Stern}
\affiliation{Jet Propulsion Laboratory, California Institute of Technology, 4800 Oak Grove Drive, Pasadena, CA 91109, USA}

\author[0000-0001-6415-0903]{Daniel Brethauer}
\affiliation{Department of Astronomy, University of California, Berkeley, CA 94720-3411, USA}

\author{Tien Duong}
\affiliation{Department of Astronomy, University of California, Berkeley, CA 94720-3411, USA}

\author[0000-0002-3934-2644]{Wynn Jacobson-Gal\'{a}n}
\affiliation{Department of Astronomy, University of California, Berkeley, CA 94720-3411, USA}

\author[0000-0002-2249-0595]{Natalie LeBaron}
\affiliation{Department of Astronomy, University of California, Berkeley, CA 94720-3411, USA}

\author[0000-0002-4513-3849]{David Matthews}
\affiliation{Department of Astronomy, University of California, Berkeley, CA 94720-3411, USA}

\author[0000-0001-8023-4912]{Huei Sears}
\affiliation{Center for Interdisciplinary Exploration and Research in Astrophysics (CIERA), Northwestern University, Evanston, IL 60202 USA}
\affiliation{Department of Physics and Astronomy, Northwestern University, Evanston, IL 60208, USA}

\author[0000-0001-8638-2780]{Padma Venkatraman}
\affiliation{Department of Astronomy, University of California, Berkeley, CA 94720-3411, USA}

\begin{abstract}
We report multi-wavelength observations and characterization of the ultraluminous transient AT 2021lwx (ZTF20abrbeie; aka ``Barbie'') identified in the alert stream of the Zwicky Transient Facility (ZTF) using a Recommender Engine For Intelligent Transient Tracking (REFITT) filter on the ANTARES alert broker. From a spectroscopically measured redshift of 0.995, we estimate a peak observed pseudo-bolometric luminosity of log\,(L$_{\text{max}} / [\text{erg}/\text{s}]$) = 45.7 from slowly fading ztf-$\it{g}$ and ztf-$r$ light curves spanning over 1000 observer-frame days. The host galaxy is not detected in archival Pan-STARRS observations ($g > 23.3$ mag), implying a lower limit to the outburst amplitude of more than 5 mag relative to the quiescent host galaxy. Optical spectra exhibit strong emission lines with narrow cores from the H Balmer series and ultraviolet semi-forbidden lines of \ion{Si}{3}] $\lambda$1892,  \ion{C}{3}] $\lambda$1909, and \ion{C}{2}] $\lambda$2325. Typical nebular lines in AGN spectra from ions such as [\ion{O}{2}] and [\ion{O}{3}] are not detected. These spectral features, along with the smooth light curve that is unlike most AGN flaring activity, and the luminosity that exceeds any observed or theorized supernova, lead us to conclude that AT 2021lwx is most likely an extreme tidal disruption event (TDE). Modeling of ZTF photometry with \texttt{MOSFiT} suggests that the TDE was between a $\approx 14 M_{\odot}$ star and a supermassive black hole of mass $M_{\text{BH}} \sim$ $10^{8} M_{\odot}$. Continued monitoring of the still-evolving light curve along with deep imaging of the field once AT\,2021lwx has faded can test this hypothesis and potentially detect the host galaxy.  
\end{abstract}


\section{Introduction} \label{sec:intro}

Recent advances in untargeted all-sky surveys have led to many new discoveries of astronomical transients related to the extreme physical conditions found in  the centers of galaxies. These discoveries have enabled transformative progress to be made in our understanding of the disruptions of stars due to tidal forces when they pass close to supermassive black holes (SMBH) called tidal disruption events (TDE; \citealt{Rees1988tde,Evans1989,Gezari2012Nature,Brown2017ASASSN14li, Gezari2021TDE}); changing low-ionization nuclear emission-line regions (LINERs; \citealt{Gezari2017Rapid,Yan2019Rapid,Neustadt2020ASASSN18jd,Frederick2019LINER}); changes/flares from existing active galactic nuclei (AGN; \citealt{Bianchi2005,Drake2011,Denney2014, Shappee2014Outburst,Frederick2021NL-Seyfert}); as well as other ambiguous nuclear transients (ANTs; \citealt{Trakhtenbrot2019, Ricci2020, Yu2022, Holoien2022asassn-17jz,Hinkle2022ASASSN-20hx}). 

\begin{figure*}[ht]
	\centering
	\includegraphics[width=\textwidth]{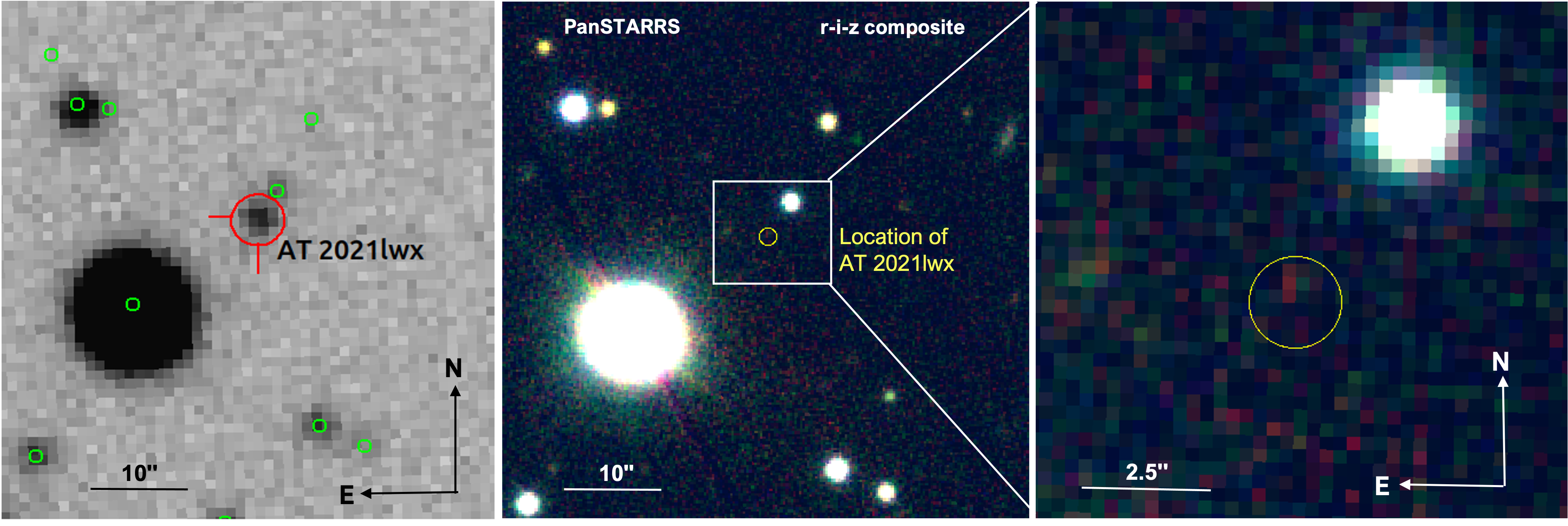}
    \caption{Left: ZTF image of the field surrounding AT 2021lwx in ztf-$r$ band obtained on 2020-12-16 and retrieved from the Infrared Science Archive (IRSA). The red circle shows the position of AT 2021lwx. Green circles are catalog stars obtained from Gaia DR2 \citep{GaiaDR22018}. Middle: Composite $r-i-z$ image of the same field made from images obtained from the PanSTARRS archive. Right: Zoom in at the location of AT 2021lwx. No underlying host galaxy is detected. }
    \label{fig:barbie_field}
\end{figure*}

Classification between these types of transient phenomena is sometimes inconclusive. TDEs in general exhibit a blue continuum with their spectra mostly dominated with broad H and He lines \citep{Gezari2012Nature,Arcavi2014,Holoien2014ASASSN14ae}. Their light curves show smooth photometric evolution with a monotonic decline that generally follow a $t^{-5/3}$ power law with timescales varying from a few days to over a year \citep{Weavers2017,Nicoll2020,Velzen2021}. Some TDEs show different decline rates and/or exhibit O III and N III optical emission lines known to originate from Bowen fluorescence (BF; \citealt{Blagorodnova2017,Blagorodnova2019, Leloudas2019,Velzen2021}). 

AGNs, in contrast, are well-studied astronomical targets showing random fluctuations including re-brightening in their light curves over a broad range of duration \citep{Angione1972,Oknyanskij1978,Bauer2009,Smith2018}. Typical AGN spectra include strong Mg II emission, relatively narrow Balmer lines and strong [O~III] emission lines in the optical \citep{Vanden2001,Aoki2005, Batra2014,Schmidt2018}. \textcolor{black}{AGN's are known to exhibit flares or outbursts of intense radiation across the electromagnetic spectrum, from radio waves to gamma rays \citep{Trakhtenbrot2019, Frederick2021NL-Seyfert}. AGN flares found in narrow-line Seyfert 1 (NLSy1) host galaxies, show narrow  Balmer as well as helium emission lines} \citep{Frederick2021NL-Seyfert}.

The scenario becomes complex when an AGN hosts a TDE, leading to a blend of features from the disruption along with its own properties \citep{Blanchard16dtm,Neustadt2020ASASSN18jd, Holoien2022asassn-17jz}. In some cases, optical TDEs show relativistic jets \citep{Zauderer2011, Burrows2011, Bloom2011,Pasham2015, Andreoni2022AT2022cmc} and may also be impacted by the spin and mass of the SMBH, making the distinction difficult \citep{GAFTON2019}.

In this paper, we report optical, ultraviolet, X-ray, and radio observations of an extremely energetic, slow evolving transient AT 2021lwx. The paper is organized as follows: Section \ref{sec:observations} describes the discovery of AT 2021lwx, along with observations that we analyze in Sections \ref{sec:properties}  and \ref{sec:spectra analysis}. We provide details about modeling AT 2021lwx as a TDE using \texttt{MOSFiT} in Section \ref{sec:light-curve-models}, and explain why AT 2021lwx is not well described by other types of transients in Section \ref{sec:comparison}. Constraints on the possible host galaxy are provided in Section \ref{sec:hostless}. Finally, in Sections \ref{sec:discussion} and \ref{sec:conclusions}, we summarize our results and conclude that AT 2021lwx is most likely a TDE with extreme properties.  

\section{Observations} \label{sec:observations}

\subsection{Optical Discovery}\label{subsec:distance}

AT 2021lwx was first identified by our group as a transient of interest in the  Zwicky Transient Facility (ZTF; \citealt{Bellm2019}) alert stream by the \texttt{refitt\_newsources\_snrcut} filter maintained by the Recommender Engine For Intelligent Transient Tracking (REFITT; \citealt{Sravan2020}; Milisavljevic et al., in preparation)\footnote{https://refitt.physics.purdue.edu/} on the ANTARES real-time alert broker \citep{Matheson21}. The filter selects objects with a signal-to-noise ratio greater than five in both ztf-$g$ and ztf-$r$ passbands, and that are located more than one arcsecond away from previously catalogued sources (\texttt{distnr} $ > 1$). Additional local processing downstream of ANTARES is done to remove false detections (e.g., poor image subtraction) and to prioritize events with light curves exhibiting features consistent with different scientific objectives. The first REFITT forecast of AT 2021lwx was on 2021-10-17. Later, as a result of adding a new classification method adopted from \citet{Garretson21}, on 2022-05-19 AT 2021lwx was flagged as a high priority Type IIn / superluminous supernova (SLSN) candidate and was triggered for follow up spectroscopic observations.  

The first discovery report of the transient to TNS was made by the Asteroid Terrestrial-impact Last Alert System (ATLAS; \citealt{Tonry2018, Smith2020}) as ATLAS20bkdj on 2020-11-10, followed by ZTF as ZTF20abrbeie on 2021-04-13. We adopt the coordinates $\alpha = 21^{\text{h}}13^{\text{m}}48.41^{\text{s}}$ and $\delta = +27^{\circ}25^{\arcmin}50.38^{\arcsec}$ (J200.0) reported by ZTF. We nicknamed the transient ``Barbie'' internally, and during the course of our monitoring campaign it was officially designated AT 2021lwx and classified as an AGN \citep{BarbieTNS} at \textit{z} = $0.995$ on 2022-09-09 by the extended Public European Southern Observatory (ESO) Spectroscopic Survey of Transient Objects (ePESSTO+; \citealt{Smartt2015}). 

\subsection{Optical\textcolor{black}{-UV} Photometry}\label{subsec:photometry}

All available ZTF public optical photometry of AT 2021lwx in ztf-$g$ and ztf-$r$ passbands was retrieved using the forced-photometry service \citep{IRSA539}. Point spread function (PSF)-fit photometry is performed on these difference images resulting in precise flux measurements \citep{Masci2019}. AT 2021lwx was also observed by the ATLAS survey in cyan and orange passbands. The data was retrieved from the ATLAS Forced Photometry server\footnote{https://fallingstar-data.com/forcedphot/}. \textcolor{black}{All ATLAS data points with both positive and negative fluxes are included in our analysis. We filter and retain only those measurements whose reduced chi-square PSF fit is between 0.5 and 5. We stack measurements in bins of seven days in each ATLAS passband to clean and reduce the scatter in the data.} 

\textcolor{black}{Two epochs of Swift - Ultraviolet Optical Telescope (UVOT; \citealt{Roming05}) photometry were acquired (PI Wang). UVOT data were reduced following standard standard prescriptions by \cite{Brown09}. A source region of 5$\arcsec$ was used for all filters. The background was estimated from a source-free region. We note that the proximity of a bright source in the $v$-band and $b$-band images leads to an increased background emission at the location of our transient of interest (and hence to a reduced sensitivity of these observations).}All magnitudes are quoted in AB magnitude system.

\textcolor{black}{From a spectroscopically measured redshift of \textit{z} = 0.995}, we adopt a distance \textit{D} $= 6.6$ Gpc and a distance modulus of $\mu = 44.09$ mag assuming a standard flat $\Lambda$CDM cosmology model with $H_{0} = 70 \: \text{km} \: \: \text{Mpc}^{-1}\: \text{s}^{-1}$ and $\Omega_{0} = 0.3$. No host galaxy associated with AT 2021lwx is visible either in any ztf-$g$ or ztf-$r$ template images,  or any band of archival Pan-STARRS images \citep[][Section \ref{sec:hostless}]{Flewelling2020PS}. Figure \ref{fig:barbie_field} shows the ZTF detection on December 16, 2020 and the corresponding field of AT 2021lwx.

\begin{figure*}[tp]
	\centering
	\includegraphics[width=\textwidth]{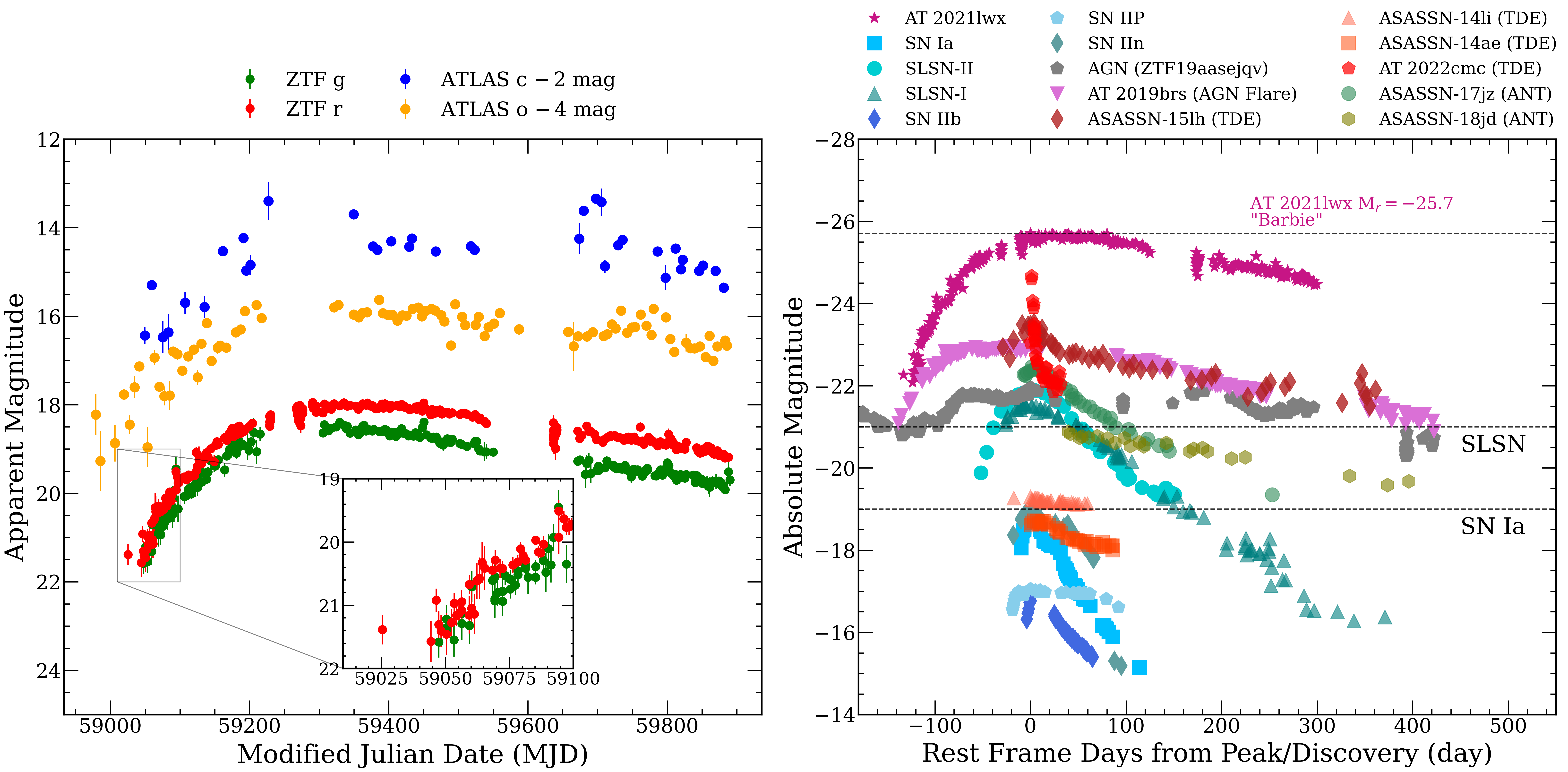}
    \caption{Optical photometry of AT 2021lwx. Left: ZTF forced photometry in ztf-$g$ and ztf-$r$ bands from the ZTF survey. The data in orange and cyan bands are from ATLAS forced photometry. The inset highlights the evidence of minor fluctuations in the otherwise smooth light curve. The light curve presented is as measured in the observed frame and the measurements are not corrected for extinction. The ATLAS photometry is plotted using an offset for the purpose of clarity. Right: Rest frame ztf-$r$ absolute light curve of AT 2021lwx plotted in comparison with other transients. The light curve of AT 2021lwx is corrected for Milky Way foreground extinction in addition to K-correction. The light curves of other transients except ASASSN-15lh (\textit{V} band) are in \textit{r/R} passbands plotted with respect to the days from peak or days from discovery (see Section \ref{sec:comparison} for references)}.
    \label{fig:photometry}
\end{figure*}

All ZTF forced photometry measurements in each passband were corrected for Milky Way extinction of $E( B -V)_{\text{MW}} = 0.12$ mag, calculated using \citet{Schlafly2011}. Figure \ref{fig:photometry} shows the optical photometry obtained in all available passbands from the ZTF and ATLAS all-sky surveys in the observed frame of reference. The rest frame light curves for AT 2021lwx were obtained after incorporating distance modulus, Milky Way reddening, time dilation and K-corrections. The observed phases were corrected by a factor of (1 + \textit{z}) and the absolute magnitudes were calculated using $\mu = 44.09$, in addition to the K-correction given by $-2.5 \text{log}\,(1 + \textit{z})$. We summarize the basic observational parameters of AT 2021lwx in Table 1. We show the optical light curves of AT 2021lwx from ZTF and ATLAS in their respective passbands in observed frame of reference in the left panel of Figure \ref{fig:photometry}. 

\input{AT2021lwx_Observed_Parameters.tex}

\begin{figure*}[tp]
	\centering
	\includegraphics[width=\textwidth]{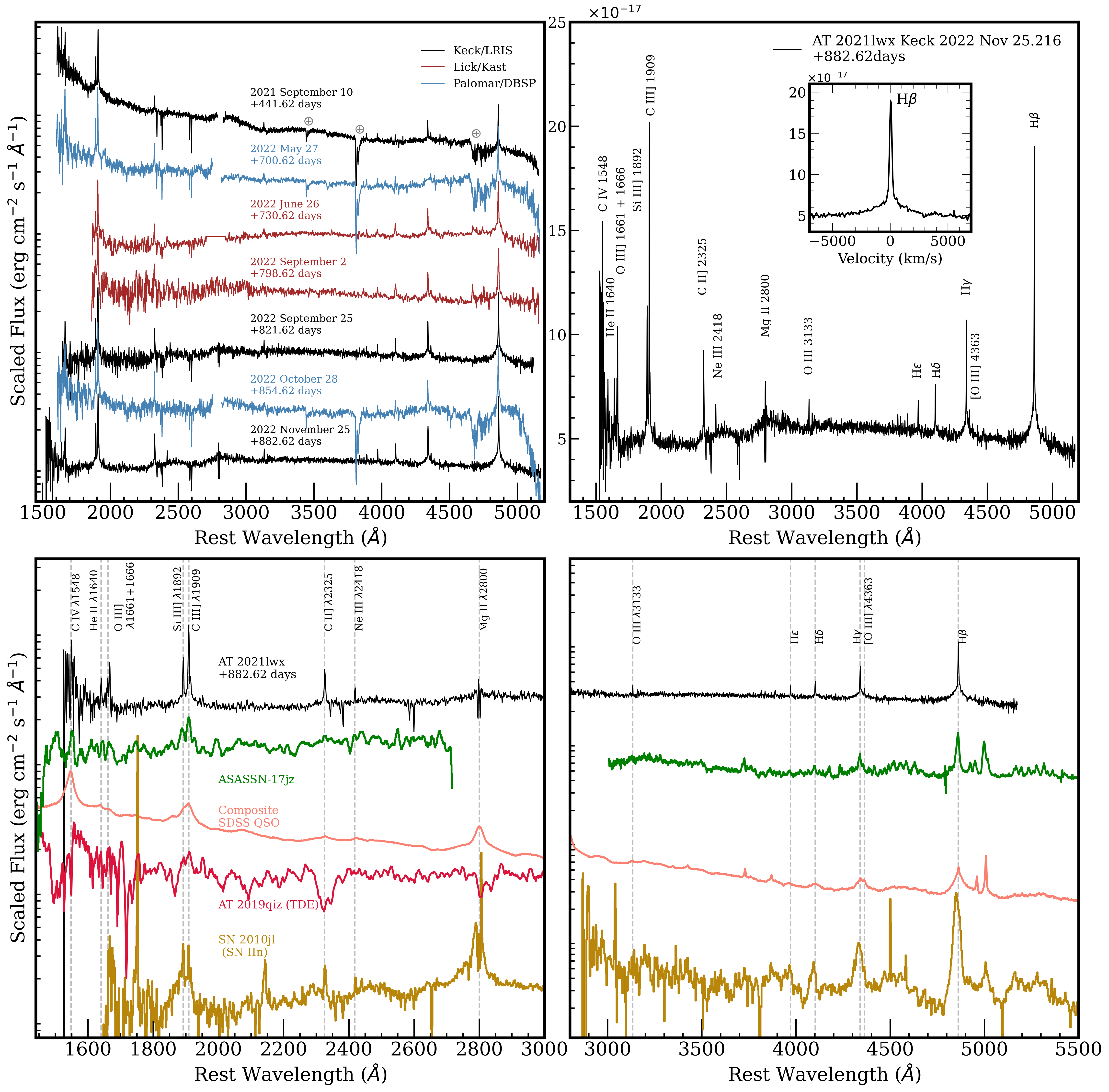}
    \caption{Top: Spectra of AT 2021lwx obtained from Lick (brown), Keck (black) and Palomar (blue) with UT dates of observations and phase with respect to the first detection in ztf-$r$ (MJD $59025.38$) are provided. Our highest quality spectrum from November 2022 is enlarged with line identifications. Bottom: Comparison with various transients including a composite quasar \citep[QSO,][]{Vanden2001}, the TDE \citep[AT 2019qiz,][]{Hung2021}, the Type IIn \citep[SN 2010jl,][]{SN2010jl2014}, and the ANT \citep[ASSASN-17jz,][]{Holoien2022asassn-17jz}. Wavelength regions of telluric contamination are marked with the ``$\Earth$'' symbol. }
    \label{fig:AT2021lwx_optical_spectra}
\end{figure*}

\subsection{Optical Spectroscopy}\label{subsec:spectroscopy}

\textcolor{black}{We obtained seven epochs of low resolution spectroscopic observations using the Low Resolution Imaging Spectrometer (LRIS; \citealt{Oke95}) on the 10~m Keck I telescope at Keck Observatory, the Kast Double Spectrograph \citep{miller93} on the Shane 3~m telescope at Lick Observatory, and the Double Spectrograph (DBSP; \citealt{Oke_1982}) mounted on the Palomar 5.1 m telescope. These spectra are shown in Figure~\ref{fig:AT2021lwx_optical_spectra}, along with UT dates of observations and observer-frame days since first detection in ztf-$r$ passband.}

Both the Kast and LRIS observations used a dichroic beamsplitter to separate the blue and red halves of the spectra. The Kast observations used the 600/4310 grism on the blue side and the 300/7500 grating on the red side, while the LRIS observations used the 400/3410 grism and 400/8500 grating to cover the full optical range. In each case, the total exposure times were slightly different on the red and blue sides because the exposures on the red side were split into a larger number of images with shorter individual exposure times, since the thick red detectors suffer from an enhanced cosmic ray rate. 

We use standard IRAF tasks to apply flat field corrections, extract one-dimensional spectra, and derive a wavelength calibration from emission-line lamps. Custom IDL tasks were used to apply a flux calibration and, when possible, correct for telluric absorption by comparison to observations of spectra of standard stars taken at an airmass comparable to that of the science target, and observed at the parallactic angle to minimize the effects of atmospheric dispersion.  

Notably, portions of the spectra of AT~2021lwx obtained on 2022 June 26 have been excised around an unreliable region between 5400~\AA\,\,and 5680~\AA, in the overlap between the blue and red sides of Kast. These data were contaminated by a nearby bright star located southeast of the transient that was intercepted by the slit and that introduced several reflections off the dichroic beamsplitter onto the object trace. 

\subsection{X-rays}\label{subsec:xray}
X-ray observations of AT 2021lwx  were obtained on two epochs (i.e. on 2022 December 10 and 2023 January 22; PI: Wang)  with the X-Ray Telescope (XRT, \citealt{Gehrels04}) on board the Neil Gehrels Observatory (total exposure time of 5.3\,ks). We reduced the data following standard practice (e.g., \citealt{Margutti13}) with HEASoft v6.31 and corresponding calibration files. We find evidence for a faint source of X-ray emission at the location of the optical transient with significance of $\approx\,3\sigma$ (targeted detection). The inferred 0.3--10 keV net count-rate is $(2.2\pm 0.87)\times 10^{-3}\,\rm{c\,s^{-1}}$. The Galactic neutral hydrogen column density in the direction of AT 2021lwx is $9.64\times 10^{20}\,\rm{cm^{-2}}$  \citep{Kalberla05}. For an assumed spectrum $F_{\nu}\propto \nu^{-1}$, the count-rate above corresponds to an unabsorbed 0.3--10 keV flux of $F_x\approx 2.1\times 10^{-13}\,\rm{erg\,s^{-1}cm^{-2}}$ (observed flux of $F_x\approx 1.1\times 10^{-13}\,\rm{erg\,s^{-1}cm^{-2}}$), which is a luminosity $L_x\approx 10^{45}\,\rm{erg\,s^{-1}}$. Within astronomical transient phenomena, this level of X-ray luminosity at late times (483 days since discovery, rest frame) has only been observationally associated with TDEs (\citealt{Polzin22}, their figures 1 and 3). However, the relativistic TDEs that have such luminous X-ray emission at late times also do not have strong emission lines in their UV-optical spectra \citep{cenko2012,Andreoni2022AT2022cmc}, in contrast with AT~2021lwx.

\subsection{Radio}\label{subsec:radio}
The location of AT\,2021lwx was observed by the Very Large Array Sky Survey (VLASS, \citealt{Lacy20}) on 2019 May 7 and 2021 November 9. The first epoch corresponds to $\approx 208$ days (rest frame) before the first optical detection of the transient, while the second epoch was acquired $\approx 250$ days (rest frame) post optical detection, which is after the optical peak. We find no evidence for significant radio emission in either observation. Using a region of 20$\arcsec$ at the location of the optical transient we infer a 3$\sigma$ upper limit on the flux density of the transient $F_{\nu}<0.35\,\rm{mJy}$ at $\approx 3\,\rm{GHz}$. This translates into a luminosity density limit of $\sim 2\times 10^{31}\,\rm{erg\,s^{-1}Hz^{-1}}$, which is a factor $\approx 10$ less luminous than on-axis jetted TDEs at the same epoch (e.g., Swift1644; \citealt{Zauderer2011, Alexander20}).

\begin{figure*}[ht]
\centering
\includegraphics[width=\linewidth]{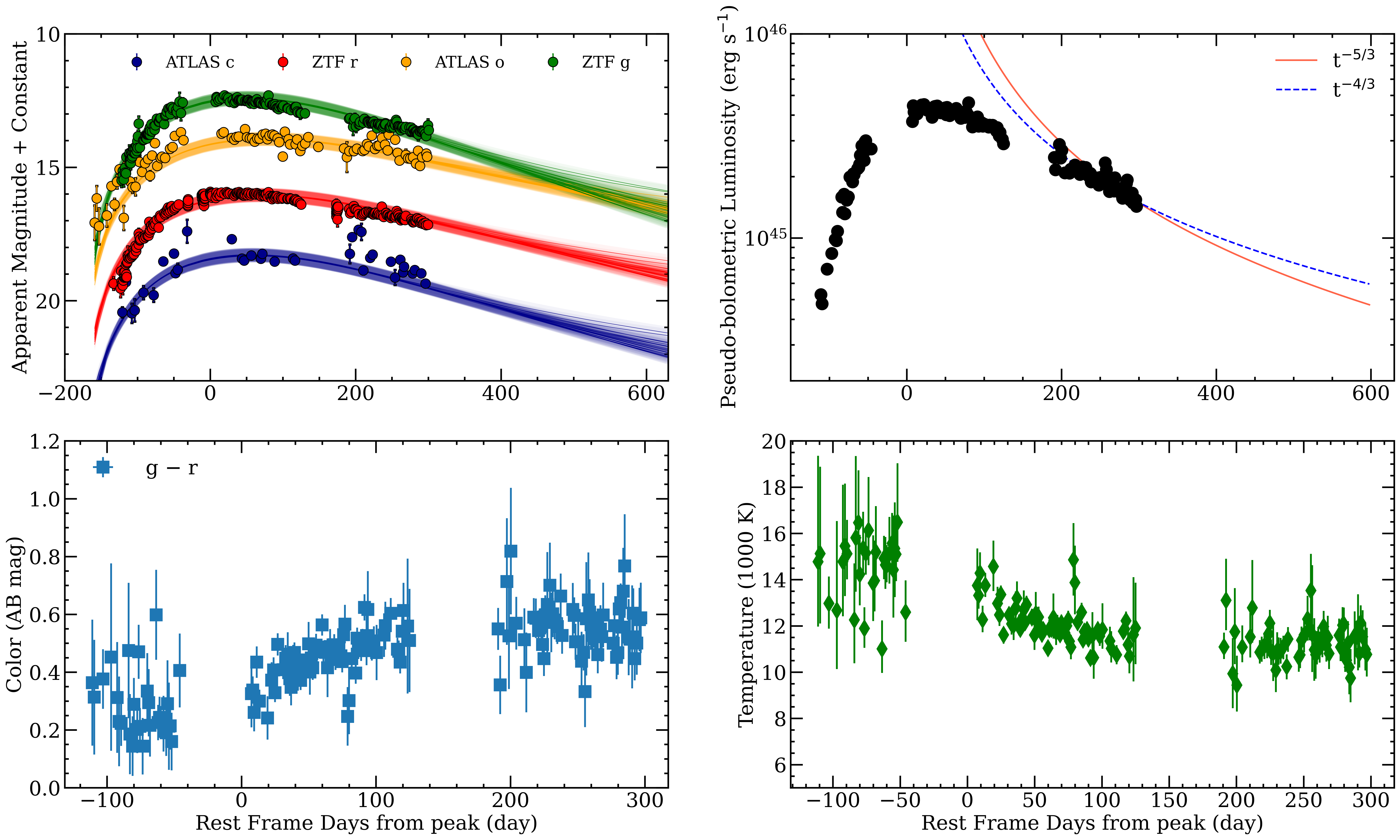}
    \caption{Top Panel: (Left) Multi-band light curve fits from MOSFiT TDE model. (Right) Pseudo-bolometric light curve of AT 2021lwx. Later epochs of AT 2021lwx are reasonably fit with  a $t^{-5/3}$ or  a $t^{-4/3}$ power law. Bottom Panel: (Left) $g$ - $r$ color evolution of AT 2021lwx. (Right) Black body temperature evolution of AT 2021lwx.}
    \label{fig:bolometric_barbie}
\end{figure*}

\section{Pseudo-bolometric Light Curve }\label{sec:properties}
\subsection{Properties}\label{subsec:pseudo-bol}

We used extinction and K-corrected measurements in ZTF passbands to construct the pseudo-bolometric light curve of AT 2021lwx. We interpolate the light curves using higher order polynomials to account for measurements at missing epochs. Each observed epoch is then fit to a black body spectral energy distribution, thereby calculating pseudo-bolometric luminosities, black body temperatures and black body radii as a function of time. We estimate the flux outside the observed passbands by extrapolating the black body fit. We only used the ZTF forced photometry measurements to construct the pseudo-bolometric light curves due to its dense sampling and long-term coverage, as compared to the ATLAS measurements. The photometry is not corrected for intrinsic host extinction, which is unknown.

AT\,2021lwx has a peak luminosity of log\,(L$_{\text{max}} / [\text{erg}/\text{s}]$) = 45.7,  making it among the most luminous transients ever observed. We measure a peak absolute magnitude of M$_{r} = - 25.7 $ mag in ztf-$r$ passband. We estimate the radiated energy of AT2021lwx to be $ E = 9.7 \times 10^{52}$ erg by integrating the bolometric light curve from -133 to 300 days in the rest-frame. We note that AT 2021lwx is still an evolving transient and the estimated energy is a lower limit. The black body fits indicate a peak temperature of 1.6 $\times 10^{4}$ K during the initial phase of evolution and subsequently drops to 1.2 $\times 10^{4}$ K. \textcolor{black}{This peak temperature of is relatively cooler compared to other TDEs with temperatures $\geq 2 \times 10^{4}$ K \citep{Velzen2021}}. The black body radius expands steadily to 3 $\times 10^{15}$ cm until approximately 90 days from peak and slowly recedes to  $10^{15}$ cm at later epochs. Figure \ref{fig:bolometric_barbie} shows the bolometric light curve, color evolution, and black body temperature evolution of AT 2021lwx. The color evolution from bluer to redder magnitudes of AT 2021lwx is consistent with the black body temperature evolution. 

\subsection{Caveats with Black Body Approximation}\label{subsec:caveats}

We note that there are limitations in using black-body approximations using data in limited passbands for accreting systems. Complex non-thermal processes, such as Compton scattering and synchrotron emission, can modify the spectrum in ways that cannot be accounted for by the black body approximation \citep{M1988,C1995,Z2020}. The assumption that the accretion disk is in local thermodynamic equilibrium (LTE) is not always the case. The temperature of the black body emission in accreting systems can be higher than that predicted due to spectral hardening from non-LTE effects \citep{H1988,D2006,D2019}. The black body approximation assumes that the emission from the accretion disk to be isotropic, however, in cases the emission could be beamed in the direction of the observer as relativistic jets \citep{M2009Jets,B2019Jets,C2021Jets}. 

\section{Spectroscopic Features}\label{sec:spectra analysis}

All spectra were obtained after the transient had evolved for more than 400 days in the observed frame since its first detection in ztf-$r$ passband. The most prominent features are strong and narrow H Balmer lines. UV lines around C IV $\lambda 1548$ and He II $\lambda 1640$  are also seen. Semi-forbidden transitions from \ion{O}{3}] $\lambda\lambda1661,1666$, Si~III] $\lambda1892$, C~III]  $\lambda1909$ and C~II]  $\lambda2325$ are conspicuous. Weak Mg II $\lambda2800$, O III $\lambda3133$ and [\ion{O}{3}] $\lambda4363$ emission is also observed in our highest quality spectrum obtained on day 883. Narrow absorption lines from the host are observed around Mg II $\lambda\lambda2796, 2803$ and standard Fe II interstellar lines. We use Lorentzian and Gaussian profiles to fit the full width at half maximum (FWHM) of the emission lines. At least two components are needed in most cases, especially the H Balmer emission and blend of Si~III] $\lambda$1892 + C~III] $\lambda$1909: a narrow unresolved Gaussian component with FWHM of $\approx 300$ km\,s$^{-1}$, and a second broader component having a half-width-zero-intensity wing that extends to $\approx 3000$ km\,s$^{-1}$. 

Our spectra of AT 2021lwx do not significantly evolve over all epochs of observations spaced over a 14 month period (observer frame). Notably, the spectra do not show emission lines commonly observed in AGN, including forbidden [Ne V] $\lambda3426$, [O II] $\lambda 3727$, [O III]  $\lambda\lambda 4959, 5007$, or the broad N III $\lambda 4640$ and  He II $\lambda 4686$ lines seen in some TDEs. We compare our spectra of AT 2021lwx to other transients in Section \ref{sec:comparison}.

We calculate an upper limit on the [\ion{O}{3}] $\lambda$5007 line flux to be 5.10 $\times 10^{-17}$ erg s$^{-1}$ cm$^{-2}$. A signal-to-noise (SNR) ratio $\sim$ 3 is used to find the threshold flux around the [\ion{O}{3}] $\lambda$5007 spectral line implying L$_{5007} \lesssim 2.7 \times 10^{41}$ erg s$^{-1}$. Then we infer an upper limit to the total bolometric luminosity of L$_{bol} \lesssim 9.5 \times 10^{44}$ erg s$^{-1}$
for any pre-existing AGN \citep{heckman04}. This is at least a factor of 5 below our estimated optical peak luminosity, implying an increase in the luminosity of the ionizing continuum by at least a similar factor.  

The strengths of the semi-forbidden lines with broader components, such as \ion{C}{3}] $\lambda$1909, and corresponding lack of {\it broad} forbidden line emission from [\ion{O}{3}] $\lambda$5007, is also a feature of the broad line regions of AGNs and has been used to estimate an electron density for the emitting gas of $n_e\approx10^9$~cm$^{-3}$ due to the collisional suppression of the forbidden lines (e.g., \citealt{agn2}).
The lack of the standard {\it narrow} nebular emission lines in a nuclear transient can also be a light travel time effect, as the narrow line regions of AGNs have been shown to have physical sizes of at least several parsecs (e.g., \citealt{peterson2013}), so a strong flare in the central ionizing source would not produce a corresponding increase in the $\lambda$5007 line flux for years or decades. This lack of forbidden line emission in the flare spectra was also seen in the turn-on of iPTF16bco \citep{Gezari2017Rapid}.

\section{Light Curve Modeling with MOSFiT}\label{sec:light-curve-models}

Accretion onto a SMBH is a plausible power source for the extreme luminosity of AT 2021lwx. Assuming a fiducial 10\% radiative efficiency, the estimated pseudo-bolometric output of $ E = 9.7 \times 10^{52}$ erg would require the accretion of $\sim$ 0.5 M$_{\odot}$ of material onto a SMBH, all within $\sim$400 rest-frame days. This large accretion rate, combined with the lack of evidence for pre-existing AGN activity as discussed in Section \ref{sec:spectra analysis}, motivates exploration of TDE scenarios as one of the few known mechanisms capable of supplying a SMBH with a sufficient supply of gas on these timescales. 

We model the multi-band light curves of AT 2021lwx using the Modular Open Source Fitter for Transients (\texttt{MOSFiT}) Python package \citep{Guillochon2018} that uses different powering mechanisms to model transients. The TDE model \citep{Mockler2019TDE} in \texttt{MOSFiT} uses a set of mass fallback curves for polytropic stars around supermassive black holes with a range of impact parameters $\beta$, defined as the ratio between tidal radius and pericenter radius. The impact parameters determine the extent of disruption of the star by the black hole. The model calculates the output bolometric luminosity by converting the input fallback rate into radiation with a given efficiency parameter \textit{$\epsilon$}. A viscous time delay approximates the speed of formation of the accretion disk around the black hole. The radiation is reprocessed by a dense extended photosphere that is related to the Eddington ratio (L/L$_{Edd}$) through a power law parameterized by the photospheric exponent \textit{l} (see Equation 10 of \citet{Mockler2019TDE}). The model assumes a black body SED that is convolved with ATLAS and ZTF passbands to estimate the magnitudes in each passband. Further details of the TDE model used in \texttt{MOSFiT} can be found in \citet{Mockler2019TDE}. 

We implement a Markov chain Monte Carlo (MCMC) sampler to fit the TDE model using \texttt{dynesty} \citep{Skilling2006} and ran the sampler until convergence. We use the default prior distributions as described in \citet{Mockler2019TDE} for the physical parameters in the TDE model to find best-fit parameters with 1-$\sigma$ uncertainties. The top left panel of Figure \ref{fig:bolometric_barbie} shows the multi-band fits to the observed data from the TDE model. The best fit model gives a black hole of mass of $M_{\text{BH}} = 1.7 \pm 0.1 \times 10^{8}$ M$_{\odot}$ tidally disrupting a star of $14.28_{-1.65}^{+0.67}$ M$_{\odot}$. The fits indicate a significant disruption of the star quantified by the parameter \textit{b} $=0.71_{-0.04}^{+0.03}$ which is a proxy for the impact parameter $\beta$. \textit{b} $=0$ corresponds to minimum disruptions while full disruptions of stars in the model will yield \textit{b} $=1$. The systematic uncertainties in the physical TDE model are quantified by \citet{Mockler2019TDE} to be 0.2 dex and 0.66 dex for black hole mass and star mass respectively. These systematic errors arise predominantly from the uncertainty associated with the mass-radius relation of the disrupted star.

\section{Comparison with other transients}\label{sec:comparison}

We compare the absolute ztf-$r$ band magnitude of AT 2021lwx with other supernova types: 
hydrogen-poor super-luminous supernova-I (SLSN-I)\citep[PTF12dam,][]{Vreeswijk2017PTF12dam}, 
SLSN-II \citep[SN 2006gy,][]{Nathan2007SN2006gy}, 
SN IIP \citep[][SN 2012aw,]{DAllora2014SN2012aw}, SN IIn \citep[SN 2005cl,][]{Kiewe2012SN2005cl}, SN Ia  \citep[SN 2014J,][]{Li2019SN2014J}  and other luminous transients: AT 2022cmc \citep{Andreoni2022AT2022cmc}, ASASSN-15lh \citep{Dong2016ASASSN15lh}, ASASSN-14li \citep{Brown2017ASASSN14li}, ASASSN-17jz \citep{Holoien2022asassn-17jz}, ASASSN-18jd \citep{Neustadt2020ASASSN18jd} and ASASSN-14ae \citep{Holoien2014ASASSN14ae}, shown in Figure \ref{fig:photometry}. \textcolor{black}{We compare \textit{r/R} band absolute magnitudes for all transients except ASASSN-15lh, for which we show absolute \textit{V} band magnitudes that span peak emission.} Rest-frame light curves are plotted relative to peak for all transients except AT 2022cmc and ASASSN-18jd for which the time is relative to the first detection. We also show different absolute magnitude thresholds for SN Ia and SLSN.

AT 2021lwx is significantly more luminous than all other transients in our sample. With an absolute magnitude of M$_{r} = -25.7$ mag, the next most luminous are AT 2022cmc \citep{Andreoni2022AT2022cmc} and ASASSN-15lh \citep{Margutti2017ASASSN15lh} with the former peaking at M$_{i} = -25$ mag  and the latter at M$_{u} = -23.5$ mag as reported in \citet{Dong2016ASASSN15lh}. \textcolor{black}{The filters for ASASSN-15lh and AT 2022cmc are chosen to roughly correspond to the same rest-frame wavelength for comparison}. AT 2021lwx exhibits a slow rise of approximately 120 rest frame days strikingly different to AT 2022cmc that exhibits rapid rise and fall time-scales. The rise and fall timescales of AT 2019brs are comparable to AT 2021lwx, but 15 times dimmer. ASASSN-15lh and ASASSAN-18jd, along with AT 2019brs, exhibit a decline rate similar to AT 2021lwx, but all are again orders of magnitude dimmer. We also compare AT 2021lwx with a normal AGN light curve of ZTF19aasejqv \citep{Hodgkin2019} that shows random fluctuations over its evolution, highlighting how AT 2021lwx's smooth photometric evolution is qualitatively different from typical AGN activity. 

In Figure \ref{fig:AT2021lwx_optical_spectra}, we compare our spectra of AT\,2021lwx to various transients: a composite quasar \citep[QSO,][]{Vanden2001}, the TDE \citep[AT 2019qiz,][]{Hung2021}, the Type IIn \citep[SN 2010jl,][]{SN2010jl2014}, and the ANT \citep[ASSASN-17jz,][]{Holoien2022asassn-17jz}.  The Balmer emission line profiles seen in SN 2010jl are evidently broader than the narrow line cores seen in AT 2021lwx. In addition, SN 2010jl exhibits strong emission from the \ion{N}{3}] multiplet between 1747--1750~\AA. The strength of these \ion{N}{3}] lines relative to \ion{C}{3}] in SN 2010jl and other interacting supernovae is due to the enhanced N abundances that are a result of the CNO process in the progenitor star \citep{SN2010jl2014}. Anomalously strong N lines were also seen in the UV spectra of the TDE ASASSN-14li \citep{cenko16}, which has also been argued to be a result of CNO processing in the interior of a massive star \citep{kochanek16}. AT~2021lwx, by contrast, lacks strong emission from these \ion{N}{3}] lines. \textcolor{black}{The lack of N emission could be attributed to the young age of the disrupted star and that CNO processing has not been substantial. Alternatively, it could be because the outer N-enriched  layer was lost prior to the disruption by the SMBH.}

\textcolor{black}{AT 2021lwx has narrower H Balmer line profiles as compared to broader ones seen in typical TDEs \citep{Brown2017ASASSN14li, Holoien2014ASASSN14ae,Hung2021,Velzen2021}. However, there are exceptions to this scenario as is noted in the case of PS16dtm \citep{Blanchard16dtm} and other TDEs (AT 2019dsg; \citealt{Cannizaro_AT2019dsg}, AT 2019meg; \citealt{Velzen2021})}. No Bowen fluorescence emission lines are seen in the spectra of AT 2021lwx, unlike some TDEs. The composite QSO spectrum shares similar Balmer emission profiles but lacks narrow semi-forbidden lines (except C III]  $\lambda1909$) as seen in AT 2021lwx. The strong [O III] $\lambda\lambda 4959, 5007$ emission doublet seen in the QSO spectrum is absent in AT 2021lwx. The closest resemblance of AT 2021lwx is with the UV and optical spectra of ASASSN-17jz. The spectra of both objects show C IV $\lambda 1548$, strong semi-forbidden lines of Si III] $\lambda1892$, C III]  $\lambda1909$ and C II]  $\lambda2325$, along with relatively narrow Balmer line profiles. 

\section{Host Galaxy}\label{sec:hostless}

No host galaxy is detected at the location of AT 2021lwx, but photometric limits can be be used to constrain the host galaxy mass. The inferred mass of $10^{7.62} M_\odot$ from the Eddington limit should be correlated with the mass of the host galaxy \citep{KH13,MP13}. This can be used to calculate the galaxy bulge as well as stellar mass of the galaxy using $M-\sigma$ relationships. We use these scaling relationships to calculate a bulge mass of log\,$(\text{M}_{\text{bulge}}/\text{M}_{\odot}) = 10.3$ and galaxy stellar mass log\,$(\text{M}_{\text{bulge}}/\text{M}_{\odot}) = 10.6$ using Equation 13 and 16 of \citet{Bentz2018}. 

The Pan-STARRS 5$\sigma$ point source limiting depth is 23.3 mag in the $g$ band and 23.2 mag in the $r$ band; accounting for Galactic dust extinction (0.45 mag in the $g$ band and 0.32 mag in the $r$ band, \citealt{Schlafly2011}), this gives an upper limit on the absolute magnitude of any potential host of $-21.3$ mag in the observed-frame $g$ band and $-21.2$ mag in the observed-frame $r$ band. From this we can extrapolate an upper limit on the stellar mass of the host galaxy. 

We make use of EzGal \citep{Mancone2012Ez} to find an upper limit on the stellar mass for different stellar population models. We use the BC03 models \citep{Bruzual2003} with a Salpeter initial mass function. The models have an exponentially declining star formation history (SFH), with an e-folding time ($\tau$) of either 0.1 Gyr or 1 Gyr, and metallicities of 0.4\,$Z_\odot$ and $Z_\odot$. \textcolor{black}{We model dust extinction within the galaxy with the Calzetti dust law \citep{Calzetti2000}, assuming nebular extinction values of E(B-V) $=$ 0, 0.1, 0.25 and 0.4.} The choice of an exponentially declining SFH is motivated by the expected redshift, $z$ $\sim$ 1; at this redshift, it is unlikely for the host galaxy to have an increasing star formation rate.
\textcolor{black}{We find the upper limit on the stellar mass for each model from the upper limit on the $r$ band, assuming various formation redshifts (translating to various ages)}.

Based on the upper limits, we infer that the Pan-STARRS non-detection does not necessarily exclude a $\sim$ 10$^{10}$ $M_\odot$ host galaxy. However, if the host galaxy is this massive, it must have long since ceased star formation. In this case, the host galaxy is faint at optical wavelengths despite being massive due to a paucity of young stars, \textcolor{black}{which is seemingly at odds with the presence of a $\sim$ 14 M$_\odot$ star. A possible resolution is that the host galaxy is a heavily dust enshrouded star forming galaxy. Extremely dust-obscured sub-millimeter selected galaxies are sufficiently massive \citep[e.g.][]{Zavala2018,Dudzeviciute2021}, but these galaxies are most common at z $\sim$ 1.5 - 2.0 \citep{Casey2013,Casey2014,Zavala2018}. Far-infrared follow-up would help to confirm or deny this possibility. An additional consideration is that stellar population synthesis models treat the galaxy as a monolith, which is an approximation. In practice, it is possible for star formation to have ceased through the majority of the galaxy but still persist in localized regions of the galaxy. The region of the galactic core in particular is still poorly understood, and recent studies suggest that the compression of in-spiralling gas could trigger star formation, especially near the SMBH \citep{Bonnel2008, Zadeh2013, Martin2018Nature}.}

\textcolor{black}{From the host galaxy stellar mass obtained from Pan-STARRS non-detection, we independently calculate the mass of SMBH using galaxy-SMBH scaling relations (Equation 16 of \citet{Bentz2018}) and infer a mass of   log\,$(\text{M}_{\text{BH}}/\text{M}_{\odot}) = 6.56$ for the underlying SMBH. This estimate is lower than the SMBH mass obtained from \texttt{MOSFiT} modeling $(\text{M}_{\text{BH}} \sim 10^{8}$). It should be noted that \citet{Bentz2018} used a sample of low redshift ($0.01 < z < 0.3$) disk galaxies to find the scaling relations and that these relations are likely to evolve with non-negligible scatter at higher redshifts, possibly due to additional stellar feedback mechanisms leading to the growth of the SMBH \citep{Delvecchio2019, Catmabacak2022}.} 

Future deep imaging of the field after AT 2021lwx fades will be necessary to further constrain the existence of the host and its mass. If the host is just below the Pan-STARRS detection limit, it should be detectable from ground-based observatories. High resolution Hubble Space Telescope (\textit{HST}) or \textit{JWST} observations would be useful to pinpoint the exact location of the transient with respect to the host and will permit improved constraints on the metallicity and stellar population.

\section{Discussion}\label{sec:discussion}

\subsection{AT\,2021lwx (``Barbie'') as an extreme TDE}

Multiple lines of evidence suggest that AT 2021lwx is mostly likely an extreme TDE with a peak luminosity of $10^{45.7}$ erg\,s$^{-1}$. \citet{Frederick2021NL-Seyfert} identified a new class of AGN flares from narrow-line Seyfert 1 (NLS1) galaxies as optical transients that exhibit significant brightness variability on timescales ranging from months to years. The spectroscopic changes include an increase in the continuum emission, as well as changes in the line emission and line ratios. They classified flares from NLS1s in their sample based on the presence of Fe II, He II, and Bowen fluorescence emission lines.  However, the spectra of AT 2021lwx lack these spectral features, making it distinct from any classified group from \cite{Frederick2021NL-Seyfert}. In addition, AT 2021lwx is significantly more luminous than  known AGN flares \citep{Peterson2004,Trakhtenbrot2019,Frederick2021NL-Seyfert}.

The Eddington limit gives an estimated minimum black hole mass of $M_{\text{BH}} = 10^{7.62} M_\odot$. Most well constrained TDEs modeled using \texttt{MOSFiT} in \citet{Mockler2019TDE} show black hole masses at the lower mass end $\lesssim$ $10^{6.5} M_\odot$. Other TDEs including AT 2018hyz \citep{Gomez2020AT2018hyz} and Swift J1644+57 \citep{Zauderer2011} exhibit typical mass ranges of $M_{\text{BH}} = 10^{6.6}$-$10^{6.95} M_{\odot}$ (also see \citet{Nicoll2022}). There is a sharp decrease in inferred TDE events with high black hole masses; an example with high black hole mass is ASASSN-15lh, proposed to originate from a highly spinning black hole with $10^{8.3} M_\odot$ \citep{Dong2016ASASSN15lh, Leloudas2016,Margutti2017ASASSN15lh}. \citet{Andreoni2022AT2022cmc} puts an upper limit on a rapidly rotating black hole mass of the jetted TDE AT 2022cmc to be $\lesssim$\,\,$5 \times 10^{8} M_\odot$. However, AT 2021lwx stands distinct from both these events in terms of luminosity, spectroscopic signatures, color and light curve evolution.

The TDE model fitting indicates disruption of a massive $14.28_{-1.65}^{+0.67}$ M$_{\odot}$ star, which is an extreme case as compared to any other known TDE. Tidal disruptions of stars with such high masses are usually unexpected due to (a) the Initial Mass Function (IMF) that heavily favors low mass stars, and (b) the relatively short lifetimes of massive stars that make it difficult for the star to form and then be scattered onto a sufficiently eccentric orbit.  On the other hand, for such a massive SMBH, a massive progenitor star may be necessary for a TDE since a massive star has a larger radius and is easier to disrupt.  A sun-like star with radius $1 R_\odot$ can potentially be disrupted by a maximally spinning black hole of $10^{7.62} M_{\odot}$, but only just barely \citep{Leloudas2016,Stone2016, Huang2023}.

The photometry more than 200 rest frame days after explosion appear to follow a $t^{-5/3}$ power-law described by \citet{Rees1988tde} (Figure~\ref{fig:bolometric_barbie}). However, a $t^{-4/3}$ power-law also provides a reasonable fit during the same time frame at later epochs. Because this event is so bright, late-time follow-up can potentially better distinguish which power law is appropriate and rule out whether this is a fallback-powered event. The ztf-$g$ band magnitude in this time range can be fit with the formula
\begin{equation*}
    m = 1.81 ~{\rm ln}[ \gamma (t-t_0) ],
\end{equation*}

\noindent with $\gamma = 61.1$ days$^{-1}$ and $t_0 =$ MJD 58953 (the coefficient of $1.81$ implying a $5/3$ decay).  If this relation holds true a year later, this could provide a strong confirmation of the TDE model.  The ztf-$r$ band data follows the same trend, but with $\gamma = 41.7$ days$^{-1}$.

\subsection{Potential massive star scenarios}

The current non-detection of the host galaxy that would be required to host a SMBH motivates consideration of massive star explosion scenarios. The total emitted energy of almost 10$^{53}$~erg requires an energy source beyond standard neutrino-powered explosions.  Superluminous SNe (SLSNe) are defined as events with absolute magnitudes less than $-21$ mag  and have been observed as bright as $-23$ mag \citep{Gal-Yam12}. Their luminosities ($>7 \times 10^{43}$ erg\,s$^{-1}$)  are greater by a factor of $\sim 10$ or more compared to normal SNe, and the additional energy is believed to be due to input from i) a central magnetar with a fast initial spin;  ii) a pair-production instability explosion that synthesizes considerable amounts of radioactive isotopes including nickel yields of up to tens of solar masses \citep{HW02,Kasen11} (known as a pair instability SN; PISN); or iii) various scenarios of circumstellar interaction. 

The light curves of SLSN models powered by a magnetar presented in \citet{Kasen10} are comparable to those of AT\,2021lwx.  However, peak luminosity of the most powerful events only reach $\sim 10^{43}$ erg\,s$^{-1}$. \textcolor{black}{The energy estimated for AT 2021lwx is at least an order of magnitude grater than the maximum energy seen in SLSNe explosions \citep{GalYam2019}}. Pair-instability supernovae (PISNe) presented in \citet{Kasen11} only approach $10^{44}$\,erg\,s$^{-1}$ at peak. 

Stars with zero-age main sequence mass (ZAMS) $\sim 70\mbox{--}140\,{M}_{\odot }$ are expected to become unstable and produce a series of energetic pulses and mass ejections before finally collapsing, called pulsational pair-instability supernovae (PPISNe). PPISNe give rise to a broad range of observable phenomena where the emitted radiation results from colliding shells \citep{Woosley2017PPISNe}.  The most luminous events of \citet{Woosley2017PPISNe} emit less than $5 \times 10^{50}$ erg and only briefly exceed $10^{44}$ erg\,s$^{-1}$. However, more extreme scenarios of PPISNe may be possible. For example, in investigating ASASSN-15lh, \citet{Chatzopoulos16} modeled ejecta-circumstellar matter interaction scenarios involving a rapidly rotating pulsational pair–instability supernova progenitor where the energy peaked at $\sim 10^{45}$ erg\,s$^{-1}$.  

\section{Conclusions}\label{sec:conclusions}

We report multi-wavelength observations of the transient AT 2021lwx identified with REFITT from the alert stream of the Zwicky Transient Facility survey. Our main conclusions can be summarized as follows:

\begin{enumerate}[itemsep=0em]

\item {AT 2021lwx is an ultraluminous, long-duration ($> $ 400 days in rest frame), energetic transient. We estimate a pseudo-bolometric peak luminosity of log\,(L$_{\text{max}} / [\text{erg}/\text{s}]$) = 45.7 and a radiated energy of $9.7 \times 10^{52}$ erg. This makes AT 2021lwx one of the most energetic and luminous transient events ever observed.}

\item {The optical spectra of AT 2021lwx show prominent H Balmer lines, \ion{O}{3}] $\lambda\lambda1661,1666$, Si III] $\lambda1892$, C III]  $\lambda1909$ and C II]  $\lambda2325$, along with weak Mg II $\lambda2800$ and O III $\lambda3133$. Nebular emission lines typically observed in AGN including [\ion{O}{2}] $\lambda$3727 and [O {\sc{iii}}] $\lambda\lambda$4959, 5007 are not detected. We do not find any significant spectroscopic evolution of AT 2021lwx over multiple epochs of observations.}

\item {There is no detection of a host galaxy in archival Pan-STARRS images covering the location of AT 2021lwx. We infer that the Pan-STARRS non-detections in $g$ and $r$ filters do not necessarily exclude a $\sim$ 10$^{10}$ $M_\odot$ host galaxy. }

\item The upper limit on [\ion{O}{3}] $\lambda$5007 emission constrains the bolometric luminosity of any pre-existing AGN. The high peak luminosity, large ($>4$~mag) increase in optical brightness, and smooth light curve are unlike normal AGN variability.

\item {We conclude that AT 2021lwx is most likely a TDE. Modeling ZTF photometry with \texttt{MOSFiT} suggests that the TDE involved a $14.28_{-1.65}^{+0.67}$ M$_{\odot}$ star and a SMBH with mass $ M_{\text{BH}} = 1.7 \pm 0.1 \times 10^{8}$ M$_{\odot}$.}

\end{enumerate}

Deep imaging of the location of AT 2021lwx once it has faded can better constrain the presence of a host galaxy. If a host galaxy is detected and our favored progenitor scenario of a TDE is correct, high resolution imaging using \textit{HST} and \textit{JWST} can determine the location of AT 2021lwx relative to the galaxy center. Follow-up observations at X-ray and radio wavelengths can also potentially provide more constraints on the underlying nature and the physical mechanisms causing AT 2021lwx.

\section*{Acknowledgements}

During the refereeing process of our manuscript, \citet{Wiseman2023} posted a manuscript to the arXiv pre-print server, which also presents a focused investigation of AT\,2021lwx. Much of their analysis agrees with our own, although in the most recent available version, they favor accretion of a giant molecular cloud by a dormant black hole of $10^8$ to $10^9$\,M$_{\odot}$.

We thank the referee for suggestions that have improved the content and presentation of this paper. D.~M.\ acknowledges NSF support from grants PHY-1914448, PHY- 2209451, AST-2037297, and AST-2206532. The TReX group at Berkeley is partially supported by NSF grants AST-2221789 and AST-2224255. W.J-G is supported by the National Science Foundation Graduate Research Fellowship Program under Grant No.~DGE-1842165. W.J-G acknowledges support from NASA grants through {\it Hubble Space Telescope} programs GO-16075, 16500 and 16922. Some of the data presented herein were obtained at the W. M. Keck Observatory, which is operated as a scientific partnership among the California Institute of Technology, the University of California and the National Aeronautics and Space Administration. The Observatory was made possible by the generous financial support of the W. M. Keck Foundation. The authors wish to recognize and acknowledge the very significant cultural role and reverence that the summit of Maunakea has always had within the indigenous Hawaiian community.  We are most fortunate to have the opportunity to conduct observations from this mountain.
A major upgrade of the Kast spectrograph on the Shane 3 m telescope at Lick Observatory, led by Brad Holden, was made possible through gifts from the Heising-Simons Foundation, William and Marina Kast, and the University of California Observatories. Research at Lick Observatory is partially supported by a generous gift from Google. We thank the staffs at Lick and Keck Observatories for their expert assistance in obtaining these observations.

This research has made use of the CIRADA cutout service at URL cutouts.cirada.ca, operated by the Canadian Initiative for Radio Astronomy Data Analysis (CIRADA). CIRADA is funded by a grant from the Canada Foundation for Innovation 2017 Innovation Fund (Project 35999), as well as by the Provinces of Ontario, British Columbia, Alberta, Manitoba and Quebec, in collaboration with the National Research Council of Canada, the US National Radio Astronomy Observatory and Australia’s Commonwealth Scientific and Industrial Research Organisation.

\facilities{Keck:I (LRIS), Shane (Kast Double spectrograph)}
\software{\texttt{MOSFiT} \citep{Guillochon2018}, PyRAF \citep{Pyraf2012}, astropy \citep{Robitaille2013astropy, Price-Whelan2018astropy} }

\bibliography{references}
\bibliographystyle{aasjournal}

\end{document}

%% file: AT2021lwx_Observed_Parameters.tex
\begin{deluxetable}{lr}[tp]
\linespread{1.0}
\setlength{\tabcolsep}{0pt} 
\tablecaption{Observational parameters of AT 2021lwx.  \label{tab:tab1}}
\tablecolumns{2}
\tablehead{}
\startdata
Right Ascension (J2000) & $21^{\text{h}}13^{\text{m}}48.41^{\text{s}}$\\
Declination (J2000) & $+27^{\circ}25^{\arcmin}50.38^{\arcsec}$ \\
Redshift (\textit{z}) & 0.995\\
Discovery Date & 2021 April 13 \\
Discovery Magnitude (ztf-$r$ band) & $18.05 \pm 0.064$ mag \\
First Detection MJD (ztf-$r$ band) & $59025.38$\\
$E( B -V)_{\text{MW}}$ &  $0.12 \pm 0.0025$ mag\\
Distance Modulus ($\mu$) & 44.09 mag \\
Peak Absolute Magnitude ($\text{M}_{abs}$) in ztf-$r$ band & $-25.7 $ mag
\enddata
\tablecomments{The observed discovery magnitude reported in ztf-$r$ band is not extinction corrected.}
\end{deluxetable}

%% file: main.bbl
\begin{thebibliography}{}
\expandafter\ifx\csname natexlab\endcsname\relax\def\natexlab#1{#1}\fi
\providecommand{\url}[1]{\href{#1}{#1}}
\providecommand{\dodoi}[1]{doi:~\href{http://doi.org/#1}{\nolinkurl{#1}}}
\providecommand{\doeprint}[1]{\href{http://ascl.net/#1}{\nolinkurl{http://ascl.net/#1}}}
\providecommand{\doarXiv}[1]{\href{https://arxiv.org/abs/#1}{\nolinkurl{https://arxiv.org/abs/#1}}}

\bibitem[{{Alexander} {et~al.}(2020){Alexander}, {van Velzen}, {Horesh}, \&
  {Zauderer}}]{Alexander20}
{Alexander}, K.~D., {van Velzen}, S., {Horesh}, A., \& {Zauderer}, B.~A. 2020,
  \ssr, 216, 81, \dodoi{10.1007/s11214-020-00702-w}

\bibitem[{{Andreoni} {et~al.}(2022){Andreoni}, {Coughlin}, {Perley}, {Yao},
  {Lu}, {Cenko}, {Kumar}, {Anand}, {Ho}, {Kasliwal}, {de Ugarte Postigo},
  {Sagues-Carracedo}, {Schulze}, {Kann}, {Kulkarni}, {Sollerman}, {Tanvir},
  {Rest}, {Izzo}, {Somalwar}, {Kaplan}, {Ahumada}, {Anupama}, {Auchettl},
  {Barway}, {Bellm}, {Bhalerao}, {Bloom}, {Bremer}, {Bulla}, {Burns},
  {Campana}, {Chandra}, {Charalampopoulos}, {Cooke}, {D'Elia}, {Kashyap Das},
  {Dobie}, {Feliciano Ag{\"u}{\'\i} Fern{\'a}ndez}, {Freeburn}, {Fremling},
  {Gezari}, {Goode}, {Graham}, {Hammerstein}, {Karambelkar}, {Kilpatrick},
  {Kool}, {Krips}, {Laher}, {Leloudas}, {Levan}, {Lundquist}, {Mahabal},
  {Medford}, {Miller}, {M{\"o}ller}, {Mooley}, {Nayana}, {Nir}, {Pang},
  {Paraskeva}, {Perley}, {Petitpas}, {Pursiainen}, {Ravi}, {Ridden-Harper},
  {Riddle}, {Rigault}, {Rodriguez}, {Rusholme}, {Sharma}, {Smith}, {Stein},
  {Th{\"o}ne}, {Tohuvavohu}, {Valdes}, {van Roestel}, {Vergani}, {Wang}, \&
  {Zhang}}]{Andreoni2022AT2022cmc}
{Andreoni}, I., {Coughlin}, M.~W., {Perley}, D.~A., {et~al.} 2022, arXiv
  e-prints, arXiv:2211.16530.
\newblock \doarXiv{2211.16530}

\bibitem[{{Angione} \& {Smith}(1972)}]{Angione1972}
{Angione}, R.~J., \& {Smith}, H.~J. 1972, in External Galaxies and
  Quasi-Stellar Objects, ed. D.~S. {Evans}, D.~{Wills}, \& B.~J. {Wills},
  Vol.~44, 171

\bibitem[{{Aoki} {et~al.}(2005){Aoki}, {Kawaguchi}, \& {Ohta}}]{Aoki2005}
{Aoki}, K., {Kawaguchi}, T., \& {Ohta}, K. 2005, \apj, 618, 601,
  \dodoi{10.1086/426075}

\bibitem[{{Arcavi} {et~al.}(2014){Arcavi}, {Gal-Yam}, {Sullivan}, {Pan},
  {Cenko}, {Horesh}, {Ofek}, {De Cia}, {Yan}, {Yang}, {Howell}, {Tal},
  {Kulkarni}, {Tendulkar}, {Tang}, {Xu}, {Sternberg}, {Cohen}, {Bloom},
  {Nugent}, {Kasliwal}, {Perley}, {Quimby}, {Miller}, {Theissen}, \&
  {Laher}}]{Arcavi2014}
{Arcavi}, I., {Gal-Yam}, A., {Sullivan}, M., {et~al.} 2014, \apj, 793, 38,
  \dodoi{10.1088/0004-637X/793/1/38}

\bibitem[{{Astropy Collaboration} {et~al.}(2013){Astropy Collaboration},
  {Robitaille}, {Tollerud}, {Greenfield}, {Droettboom}, {Bray}, {Aldcroft},
  {Davis}, {Ginsburg}, {Price-Whelan}, {Kerzendorf}, {Conley}, {Crighton},
  {Barbary}, {Muna}, {Ferguson}, {Grollier}, {Parikh}, {Nair}, {Unther},
  {Deil}, {Woillez}, {Conseil}, {Kramer}, {Turner}, {Singer}, {Fox}, {Weaver},
  {Zabalza}, {Edwards}, {Azalee Bostroem}, {Burke}, {Casey}, {Crawford},
  {Dencheva}, {Ely}, {Jenness}, {Labrie}, {Lim}, {Pierfederici}, {Pontzen},
  {Ptak}, {Refsdal}, {Servillat}, \& {Streicher}}]{Robitaille2013astropy}
{Astropy Collaboration}, {Robitaille}, T.~P., {Tollerud}, E.~J., {et~al.} 2013,
  \aap, 558, A33, \dodoi{10.1051/0004-6361/201322068}

\bibitem[{{Astropy Collaboration} {et~al.}(2018){Astropy Collaboration},
  {Price-Whelan}, {Sip{\H{o}}cz}, {G{\"u}nther}, {Lim}, {Crawford}, {Conseil},
  {Shupe}, {Craig}, {Dencheva}, {Ginsburg}, {Vand erPlas}, {Bradley},
  {P{\'e}rez-Su{\'a}rez}, {de Val-Borro}, {Aldcroft}, {Cruz}, {Robitaille},
  {Tollerud}, {Ardelean}, {Babej}, {Bach}, {Bachetti}, {Bakanov}, {Bamford},
  {Barentsen}, {Barmby}, {Baumbach}, {Berry}, {Biscani}, {Boquien}, {Bostroem},
  {Bouma}, {Brammer}, {Bray}, {Breytenbach}, {Buddelmeijer}, {Burke},
  {Calderone}, {Cano Rodr{\'\i}guez}, {Cara}, {Cardoso}, {Cheedella}, {Copin},
  {Corrales}, {Crichton}, {D'Avella}, {Deil}, {Depagne}, {Dietrich}, {Donath},
  {Droettboom}, {Earl}, {Erben}, {Fabbro}, {Ferreira}, {Finethy}, {Fox},
  {Garrison}, {Gibbons}, {Goldstein}, {Gommers}, {Greco}, {Greenfield},
  {Groener}, {Grollier}, {Hagen}, {Hirst}, {Homeier}, {Horton}, {Hosseinzadeh},
  {Hu}, {Hunkeler}, {Ivezi{\'c}}, {Jain}, {Jenness}, {Kanarek}, {Kendrew},
  {Kern}, {Kerzendorf}, {Khvalko}, {King}, {Kirkby}, {Kulkarni}, {Kumar},
  {Lee}, {Lenz}, {Littlefair}, {Ma}, {Macleod}, {Mastropietro}, {McCully},
  {Montagnac}, {Morris}, {Mueller}, {Mumford}, {Muna}, {Murphy}, {Nelson},
  {Nguyen}, {Ninan}, {N{\"o}the}, {Ogaz}, {Oh}, {Parejko}, {Parley}, {Pascual},
  {Patil}, {Patil}, {Plunkett}, {Prochaska}, {Rastogi}, {Reddy Janga},
  {Sabater}, {Sakurikar}, {Seifert}, {Sherbert}, {Sherwood-Taylor}, {Shih},
  {Sick}, {Silbiger}, {Singanamalla}, {Singer}, {Sladen}, {Sooley},
  {Sornarajah}, {Streicher}, {Teuben}, {Thomas}, {Tremblay}, {Turner},
  {Terr{\'o}n}, {van Kerkwijk}, {de la Vega}, {Watkins}, {Weaver}, {Whitmore},
  {Woillez}, {Zabalza}, \& {Astropy Contributors}}]{Price-Whelan2018astropy}
{Astropy Collaboration}, {Price-Whelan}, A.~M., {Sip{\H{o}}cz}, B.~M., {et~al.}
  2018, \aj, 156, 123, \dodoi{10.3847/1538-3881/aabc4f}

\bibitem[{{Batra} \& {Baldwin}(2014)}]{Batra2014}
{Batra}, N.~D., \& {Baldwin}, J.~A. 2014, \mnras, 439, 771,
  \dodoi{10.1093/mnras/stu007}

\bibitem[{{Bauer} {et~al.}(2009){Bauer}, {Baltay}, {Coppi}, {Ellman}, {Jerke},
  {Rabinowitz}, \& {Scalzo}}]{Bauer2009}
{Bauer}, A., {Baltay}, C., {Coppi}, P., {et~al.} 2009, \apj, 696, 1241,
  \dodoi{10.1088/0004-637X/696/2/1241}

\bibitem[{{Bellm} {et~al.}(2019){Bellm}, {Kulkarni}, {Graham}, {Dekany},
  {Smith}, {Riddle}, {Masci}, {Helou}, {Prince}, {Adams}, {Barbarino},
  {Barlow}, {Bauer}, {Beck}, {Belicki}, {Biswas}, {Blagorodnova}, {Bodewits},
  {Bolin}, {Brinnel}, {Brooke}, {Bue}, {Bulla}, {Burruss}, {Cenko}, {Chang},
  {Connolly}, {Coughlin}, {Cromer}, {Cunningham}, {De}, {Delacroix}, {Desai},
  {Duev}, {Eadie}, {Farnham}, {Feeney}, {Feindt}, {Flynn}, {Franckowiak},
  {Frederick}, {Fremling}, {Gal-Yam}, {Gezari}, {Giomi}, {Goldstein},
  {Golkhou}, {Goobar}, {Groom}, {Hacopians}, {Hale}, {Henning}, {Ho}, {Hover},
  {Howell}, {Hung}, {Huppenkothen}, {Imel}, {Ip}, {Ivezi{\'c}}, {Jackson},
  {Jones}, {Juric}, {Kasliwal}, {Kaspi}, {Kaye}, {Kelley}, {Kowalski},
  {Kramer}, {Kupfer}, {Landry}, {Laher}, {Lee}, {Lin}, {Lin}, {Lunnan},
  {Giomi}, {Mahabal}, {Mao}, {Miller}, {Monkewitz}, {Murphy}, {Ngeow},
  {Nordin}, {Nugent}, {Ofek}, {Patterson}, {Penprase}, {Porter}, {Rauch},
  {Rebbapragada}, {Reiley}, {Rigault}, {Rodriguez}, {van Roestel}, {Rusholme},
  {van Santen}, {Schulze}, {Shupe}, {Singer}, {Soumagnac}, {Stein}, {Surace},
  {Sollerman}, {Szkody}, {Taddia}, {Terek}, {Van Sistine}, {van Velzen},
  {Vestrand}, {Walters}, {Ward}, {Ye}, {Yu}, {Yan}, \& {Zolkower}}]{Bellm2019}
{Bellm}, E.~C., {Kulkarni}, S.~R., {Graham}, M.~J., {et~al.} 2019, \pasp, 131,
  018002, \dodoi{10.1088/1538-3873/aaecbe}

\bibitem[{{Bentz} \& {Manne-Nicholas}(2018)}]{Bentz2018}
{Bentz}, M.~C., \& {Manne-Nicholas}, E. 2018, \apj, 864, 146,
  \dodoi{10.3847/1538-4357/aad808}

\bibitem[{{Bianchi} {et~al.}(2005){Bianchi}, {Guainazzi}, {Matt}, {Chiaberge},
  {Iwasawa}, {Fiore}, \& {Maiolino}}]{Bianchi2005}
{Bianchi}, S., {Guainazzi}, M., {Matt}, G., {et~al.} 2005, \aap, 442, 185,
  \dodoi{10.1051/0004-6361:20053389}

\bibitem[{{Blagorodnova} {et~al.}(2017){Blagorodnova}, {Gezari}, {Hung},
  {Kulkarni}, {Cenko}, {Pasham}, {Yan}, {Arcavi}, {Ben-Ami}, {Bue}, {Cantwell},
  {Cao}, {Castro-Tirado}, {Fender}, {Fremling}, {Gal-Yam}, {Ho}, {Horesh},
  {Hosseinzadeh}, {Kasliwal}, {Kong}, {Laher}, {Leloudas}, {Lunnan}, {Masci},
  {Mooley}, {Neill}, {Nugent}, {Powell}, {Valeev}, {Vreeswijk}, {Walters}, \&
  {Wozniak}}]{Blagorodnova2017}
{Blagorodnova}, N., {Gezari}, S., {Hung}, T., {et~al.} 2017, \apj, 844, 46,
  \dodoi{10.3847/1538-4357/aa7579}

\bibitem[{{Blagorodnova} {et~al.}(2019){Blagorodnova}, {Cenko}, {Kulkarni},
  {Arcavi}, {Bloom}, {Duggan}, {Filippenko}, {Fremling}, {Horesh},
  {Hosseinzadeh}, {Karamehmetoglu}, {Levan}, {Masci}, {Nugent}, {Pasham},
  {Veilleux}, {Walters}, {Yan}, \& {Zheng}}]{Blagorodnova2019}
{Blagorodnova}, N., {Cenko}, S.~B., {Kulkarni}, S.~R., {et~al.} 2019, \apj,
  873, 92, \dodoi{10.3847/1538-4357/ab04b0}

\bibitem[{{Blanchard} {et~al.}(2017){Blanchard}, {Nicholl}, {Berger},
  {Guillochon}, {Margutti}, {Chornock}, {Alexander}, {Leja}, \&
  {Drout}}]{Blanchard16dtm}
{Blanchard}, P.~K., {Nicholl}, M., {Berger}, E., {et~al.} 2017, \apj, 843, 106,
  \dodoi{10.3847/1538-4357/aa77f7}

\bibitem[{{Blandford} {et~al.}(2019){Blandford}, {Meier}, \&
  {Readhead}}]{B2019Jets}
{Blandford}, R., {Meier}, D., \& {Readhead}, A. 2019, \araa, 57, 467,
  \dodoi{10.1146/annurev-astro-081817-051948}

\bibitem[{{Bloom} {et~al.}(2011){Bloom}, {Giannios}, {Metzger}, {Cenko},
  {Perley}, {Butler}, {Tanvir}, {Levan}, {O'Brien}, {Strubbe}, {De Colle},
  {Ramirez-Ruiz}, {Lee}, {Nayakshin}, {Quataert}, {King}, {Cucchiara},
  {Guillochon}, {Bower}, {Fruchter}, {Morgan}, \& {van der Horst}}]{Bloom2011}
{Bloom}, J.~S., {Giannios}, D., {Metzger}, B.~D., {et~al.} 2011, Science, 333,
  203, \dodoi{10.1126/science.1207150}

\bibitem[{{Bonnell} \& {Rice}(2008)}]{Bonnel2008}
{Bonnell}, I.~A., \& {Rice}, W.~K.~M. 2008, Science, 321, 1060,
  \dodoi{10.1126/science.1160653}

\bibitem[{{Brown} {et~al.}(2017){Brown}, {Holoien}, {Auchettl}, {Stanek},
  {Kochanek}, {Shappee}, {Prieto}, \& {Grupe}}]{Brown2017ASASSN14li}
{Brown}, J.~S., {Holoien}, T.~W.~S., {Auchettl}, K., {et~al.} 2017, \mnras,
  466, 4904, \dodoi{10.1093/mnras/stx033}

\bibitem[{{Brown} {et~al.}(2009){Brown}, {Holland}, {Immler}, {Milne},
  {Roming}, {Gehrels}, {Nousek}, {Panagia}, {Still}, \& {Vanden
  Berk}}]{Brown09}
{Brown}, P.~J., {Holland}, S.~T., {Immler}, S., {et~al.} 2009, \aj, 137, 4517,
  \dodoi{10.1088/0004-6256/137/5/4517}

\bibitem[{{Bruzual} \& {Charlot}(2003)}]{Bruzual2003}
{Bruzual}, G., \& {Charlot}, S. 2003, \mnras, 344, 1000,
  \dodoi{10.1046/j.1365-8711.2003.06897.x}

\bibitem[{{Burrows} {et~al.}(2011){Burrows}, {Kennea}, {Ghisellini}, {Mangano},
  {Zhang}, {Page}, {Eracleous}, {Romano}, {Sakamoto}, {Falcone}, {Osborne},
  {Campana}, {Beardmore}, {Breeveld}, {Chester}, {Corbet}, {Covino},
  {Cummings}, {D'Avanzo}, {D'Elia}, {Esposito}, {Evans}, {Fugazza}, {Gelbord},
  {Hiroi}, {Holland}, {Huang}, {Im}, {Israel}, {Jeon}, {Jeon}, {Jun}, {Kawai},
  {Kim}, {Krimm}, {Marshall}, {P. M{\'e}sz{\'a}ros}, {Negoro}, {Omodei},
  {Park}, {Perkins}, {Sugizaki}, {Sung}, {Tagliaferri}, {Troja}, {Ueda},
  {Urata}, {Usui}, {Antonelli}, {Barthelmy}, {Cusumano}, {Giommi}, {Melandri},
  {Perri}, {Racusin}, {Sbarufatti}, {Siegel}, \& {Gehrels}}]{Burrows2011}
{Burrows}, D.~N., {Kennea}, J.~A., {Ghisellini}, G., {et~al.} 2011, \nat, 476,
  421, \dodoi{10.1038/nature10374}

\bibitem[{{Calzetti} {et~al.}(2000){Calzetti}, {Armus}, {Bohlin}, {Kinney},
  {Koornneef}, \& {Storchi-Bergmann}}]{Calzetti2000}
{Calzetti}, D., {Armus}, L., {Bohlin}, R.~C., {et~al.} 2000, \apj, 533, 682,
  \dodoi{10.1086/308692}

\bibitem[{{Cannizzaro} {et~al.}(2021){Cannizzaro}, {Wevers}, {Jonker},
  {P{\'e}rez-Torres}, {Moldon}, {Mata-S{\'a}nchez}, {Leloudas}, {Pasham},
  {Mattila}, {Arcavi}, {Decker French}, {Onori}, {Inserra}, {Nicholl},
  {Gromadzki}, {Chen}, {M{\"u}ller-Bravo}, {Short}, {Anderson}, {Young},
  {Gendreau}, {Arzoumanian}, {L{\"o}wenstein}, {Remillard}, {Roy}, \&
  {Hiramatsu}}]{Cannizaro_AT2019dsg}
{Cannizzaro}, G., {Wevers}, T., {Jonker}, P.~G., {et~al.} 2021, \mnras, 504,
  792, \dodoi{10.1093/mnras/stab851}

\bibitem[{{Casey} {et~al.}(2014){Casey}, {Narayanan}, \& {Cooray}}]{Casey2014}
{Casey}, C.~M., {Narayanan}, D., \& {Cooray}, A. 2014, \physrep, 541, 45,
  \dodoi{10.1016/j.physrep.2014.02.009}

\bibitem[{{Casey} {et~al.}(2013){Casey}, {Chen}, {Cowie}, {Barger}, {Capak},
  {Ilbert}, {Koss}, {Lee}, {Le Floc'h}, {Sanders}, \& {Williams}}]{Casey2013}
{Casey}, C.~M., {Chen}, C.-C., {Cowie}, L.~L., {et~al.} 2013, \mnras, 436,
  1919, \dodoi{10.1093/mnras/stt1673}

\bibitem[{{{\c{C}}atmabacak} {et~al.}(2022){{\c{C}}atmabacak}, {Feldmann},
  {Angl{\'e}s-Alc{\'a}zar}, {Faucher-Gigu{\`e}re}, {Hopkins}, \&
  {Kere{\v{s}}}}]{Catmabacak2022}
{{\c{C}}atmabacak}, O., {Feldmann}, R., {Angl{\'e}s-Alc{\'a}zar}, D., {et~al.}
  2022, \mnras, 511, 506, \dodoi{10.1093/mnras/stac040}

\bibitem[{{Cenko} {et~al.}(2012){Cenko}, {Krimm}, {Horesh}, {Rau}, {Frail},
  {Kennea}, {Levan}, {Holland}, {Butler}, {Quimby}, {Bloom}, {Filippenko},
  {Gal-Yam}, {Greiner}, {Kulkarni}, {Ofek}, {Olivares E.}, {Schady},
  {Silverman}, {Tanvir}, \& {Xu}}]{cenko2012}
{Cenko}, S.~B., {Krimm}, H.~A., {Horesh}, A., {et~al.} 2012, \apj, 753, 77,
  \dodoi{10.1088/0004-637X/753/1/77}

\bibitem[{{Cenko} {et~al.}(2016){Cenko}, {Cucchiara}, {Roth}, {Veilleux},
  {Prochaska}, {Yan}, {Guillochon}, {Maksym}, {Arcavi}, {Butler}, {Filippenko},
  {Fruchter}, {Gezari}, {Kasen}, {Levan}, {Miller}, {Pasham}, {Ramirez-Ruiz},
  {Strubbe}, {Tanvir}, \& {Tombesi}}]{cenko16}
{Cenko}, S.~B., {Cucchiara}, A., {Roth}, N., {et~al.} 2016, \apjl, 818, L32,
  \dodoi{10.3847/2041-8205/818/2/L32}

\bibitem[{{Chakrabarti} \& {Titarchuk}(1995)}]{C1995}
{Chakrabarti}, S., \& {Titarchuk}, L.~G. 1995, \apj, 455, 623,
  \dodoi{10.1086/176610}

\bibitem[{{Chatzopoulos} {et~al.}(2016){Chatzopoulos}, {Wheeler}, {Vinko},
  {Nagy}, {Wiggins}, \& {Even}}]{Chatzopoulos16}
{Chatzopoulos}, E., {Wheeler}, J.~C., {Vinko}, J., {et~al.} 2016, \apj, 828,
  94, \dodoi{10.3847/0004-637X/828/2/94}

\bibitem[{{Chen} {et~al.}(2021){Chen}, {Gu}, {Fan}, {Zhou}, {Yuan}, {Gu}, {Wu},
  {Xiong}, {Guo}, {Ding}, \& {Yu}}]{C2021Jets}
{Chen}, Y., {Gu}, Q., {Fan}, J., {et~al.} 2021, \apj, 913, 93,
  \dodoi{10.3847/1538-4357/abf4ff}

\bibitem[{{Dall'Ora} {et~al.}(2014){Dall'Ora}, {Botticella}, {Pumo},
  {Zampieri}, {Tomasella}, {Pignata}, {Bayless}, {Pritchard}, {Taubenberger},
  {Kotak}, {Inserra}, {Della Valle}, {Cappellaro}, {Benetti}, {Benitez},
  {Bufano}, {Elias-Rosa}, {Fraser}, {Haislip}, {Harutyunyan}, {Howell},
  {Hsiao}, {Iijima}, {Kankare}, {Kuin}, {Maund}, {Morales-Garoffolo},
  {Morrell}, {Munari}, {Ochner}, {Pastorello}, {Patat}, {Phillips}, {Reichart},
  {Roming}, {Siviero}, {Smartt}, {Sollerman}, {Taddia}, {Valenti}, \&
  {Wright}}]{DAllora2014SN2012aw}
{Dall'Ora}, M., {Botticella}, M.~T., {Pumo}, M.~L., {et~al.} 2014, \apj, 787,
  139, \dodoi{10.1088/0004-637X/787/2/139}

\bibitem[{{Davis} \& {El-Abd}(2019)}]{D2019}
{Davis}, S.~W., \& {El-Abd}, S. 2019, \apj, 874, 23,
  \dodoi{10.3847/1538-4357/ab05c5}

\bibitem[{{Davis} \& {Hubeny}(2006)}]{D2006}
{Davis}, S.~W., \& {Hubeny}, I. 2006, \apjs, 164, 530, \dodoi{10.1086/503549}

\bibitem[{{Delvecchio} {et~al.}(2019){Delvecchio}, {Daddi}, {Shankar},
  {Mullaney}, {Zamorani}, {Aird}, {Bernhard}, {Cimatti}, {Elbaz}, {Giavalisco},
  \& {Grimmett}}]{Delvecchio2019}
{Delvecchio}, I., {Daddi}, E., {Shankar}, F., {et~al.} 2019, \apjl, 885, L36,
  \dodoi{10.3847/2041-8213/ab4e21}

\bibitem[{{Denney} {et~al.}(2014){Denney}, {De Rosa}, {Croxall}, {Gupta},
  {Bentz}, {Fausnaugh}, {Grier}, {Martini}, {Mathur}, {Peterson}, {Pogge}, \&
  {Shappee}}]{Denney2014}
{Denney}, K.~D., {De Rosa}, G., {Croxall}, K., {et~al.} 2014, \apj, 796, 134,
  \dodoi{10.1088/0004-637X/796/2/134}

\bibitem[{{Dong} {et~al.}(2016){Dong}, {Shappee}, {Prieto}, {Jha}, {Stanek},
  {Holoien}, {Kochanek}, {Thompson}, {Morrell}, {Thompson}, {Basu}, {Beacom},
  {Bersier}, {Brimacombe}, {Brown}, {Bufano}, {Chen}, {Conseil}, {Danilet},
  {Falco}, {Grupe}, {Kiyota}, {Masi}, {Nicholls}, {Olivares E.}, {Pignata},
  {Pojmanski}, {Simonian}, {Szczygiel}, \& {Wo{\'z}niak}}]{Dong2016ASASSN15lh}
{Dong}, S., {Shappee}, B.~J., {Prieto}, J.~L., {et~al.} 2016, Science, 351,
  257, \dodoi{10.1126/science.aac9613}

\bibitem[{{Drake} {et~al.}(2011){Drake}, {Djorgovski}, {Mahabal}, {Anderson},
  {Roy}, {Mohan}, {Ravindranath}, {Frail}, {Gezari}, {Neill}, {Ho}, {Prieto},
  {Thompson}, {Thorstensen}, {Wagner}, {Kowalski}, {Chiang}, {Grove},
  {Schinzel}, {Wood}, {Carrasco}, {Recillas}, {Kewley}, {Archana}, {Basu},
  {Wadadekar}, {Kumar}, {Myers}, {Phinney}, {Williams}, {Graham}, {Catelan},
  {Beshore}, {Larson}, \& {Christensen}}]{Drake2011}
{Drake}, A.~J., {Djorgovski}, S.~G., {Mahabal}, A., {et~al.} 2011, \apj, 735,
  106, \dodoi{10.1088/0004-637X/735/2/106}

\bibitem[{{Dudzevi{\v{c}}i{\={u}}t{\.{e}}}
  {et~al.}(2021){Dudzevi{\v{c}}i{\={u}}t{\.{e}}}, {Smail}, {Swinbank}, {Lim},
  {Wang}, {Simpson}, {Ao}, {Chapman}, {Chen}, {Clements}, {Dannerbauer}, {Ho},
  {Hwang}, {Koprowski}, {Lee}, {Scott}, {Shim}, {Shirley}, \&
  {Toba}}]{Dudzeviciute2021}
{Dudzevi{\v{c}}i{\={u}}t{\.{e}}}, U., {Smail}, I., {Swinbank}, A.~M., {et~al.}
  2021, \mnras, 500, 942, \dodoi{10.1093/mnras/staa3285}

\bibitem[{{Evans} \& {Kochanek}(1989)}]{Evans1989}
{Evans}, C.~R., \& {Kochanek}, C.~S. 1989, \apjl, 346, L13,
  \dodoi{10.1086/185567}

\bibitem[{{Flewelling} {et~al.}(2020){Flewelling}, {Magnier}, {Chambers},
  {Heasley}, {Holmberg}, {Huber}, {Sweeney}, {Waters}, {Calamida}, {Casertano},
  {Chen}, {Farrow}, {Hasinger}, {Henderson}, {Long}, {Metcalfe}, {Narayan},
  {Nieto-Santisteban}, {Norberg}, {Rest}, {Saglia}, {Szalay}, {Thakar},
  {Tonry}, {Valenti}, {Werner}, {White}, {Denneau}, {Draper}, {Hodapp},
  {Jedicke}, {Kaiser}, {Kudritzki}, {Price}, {Wainscoat}, {Chastel}, {McLean},
  {Postman}, \& {Shiao}}]{Flewelling2020PS}
{Flewelling}, H.~A., {Magnier}, E.~A., {Chambers}, K.~C., {et~al.} 2020, \apjs,
  251, 7, \dodoi{10.3847/1538-4365/abb82d}

\bibitem[{{Fransson} {et~al.}(2014){Fransson}, {Ergon}, {Challis}, {Chevalier},
  {France}, {Kirshner}, {Marion}, {Milisavljevic}, {Smith}, {Bufano},
  {Friedman}, {Kangas}, {Larsson}, {Mattila}, {Benetti}, {Chornock}, {Czekala},
  {Soderberg}, \& {Sollerman}}]{SN2010jl2014}
{Fransson}, C., {Ergon}, M., {Challis}, P.~J., {et~al.} 2014, \apj, 797, 118,
  \dodoi{10.1088/0004-637X/797/2/118}

\bibitem[{{Frederick} {et~al.}(2019){Frederick}, {Gezari}, {Graham}, {Cenko},
  {van Velzen}, {Stern}, {Blagorodnova}, {Kulkarni}, {Yan}, {De}, {Fremling},
  {Hung}, {Kara}, {Shupe}, {Ward}, {Bellm}, {Dekany}, {Duev}, {Feindt},
  {Giomi}, {Kupfer}, {Laher}, {Masci}, {Miller}, {Neill}, {Ngeow}, {Patterson},
  {Porter}, {Rusholme}, {Sollerman}, \& {Walters}}]{Frederick2019LINER}
{Frederick}, S., {Gezari}, S., {Graham}, M.~J., {et~al.} 2019, \apj, 883, 31,
  \dodoi{10.3847/1538-4357/ab3a38}

\bibitem[{{Frederick} {et~al.}(2021){Frederick}, {Gezari}, {Graham},
  {Sollerman}, {van Velzen}, {Perley}, {Stern}, {Ward}, {Hammerstein}, {Hung},
  {Yan}, {Andreoni}, {Bellm}, {Duev}, {Kowalski}, {Mahabal}, {Masci},
  {Medford}, {Rusholme}, {Smith}, \& {Walters}}]{Frederick2021NL-Seyfert}
---. 2021, \apj, 920, 56, \dodoi{10.3847/1538-4357/ac110f}

\bibitem[{{Gafton} \& {Rosswog}(2019)}]{GAFTON2019}
{Gafton}, E., \& {Rosswog}, S. 2019, \mnras, 487, 4790,
  \dodoi{10.1093/mnras/stz1530}

\bibitem[{{Gal-Yam}(2012)}]{Gal-Yam12}
{Gal-Yam}, A. 2012, Science, 337, 927, \dodoi{10.1126/science.1203601}

\bibitem[{{Gal-Yam}(2019)}]{GalYam2019}
---. 2019, \araa, 57, 305, \dodoi{10.1146/annurev-astro-081817-051819}

\bibitem[{{Garretson} {et~al.}(2021){Garretson}, {Milisavljevic}, {Reynolds},
  {Weil}, {Subrayan}, {Banovetz}, \& {Lee}}]{Garretson21}
{Garretson}, B., {Milisavljevic}, D., {Reynolds}, J., {et~al.} 2021, Research
  Notes of the American Astronomical Society, 5, 283,
  \dodoi{10.3847/2515-5172/ac416e}

\bibitem[{{Gehrels} {et~al.}(2004){Gehrels}, {Chincarini}, {Giommi}, {Mason},
  {Nousek}, {Wells}, {White}, {Barthelmy}, {Burrows}, {Cominsky}, {Hurley},
  {Marshall}, {M{\'e}sz{\'a}ros}, {Roming}, {Angelini}, {Barbier}, {Belloni},
  {Campana}, {Caraveo}, {Chester}, {Citterio}, {Cline}, {Cropper}, {Cummings},
  {Dean}, {Feigelson}, {Fenimore}, {Frail}, {Fruchter}, {Garmire}, {Gendreau},
  {Ghisellini}, {Greiner}, {Hill}, {Hunsberger}, {Krimm}, {Kulkarni}, {Kumar},
  {Lebrun}, {Lloyd-Ronning}, {Markwardt}, {Mattson}, {Mushotzky}, {Norris},
  {Osborne}, {Paczynski}, {Palmer}, {Park}, {Parsons}, {Paul}, {Rees},
  {Reynolds}, {Rhoads}, {Sasseen}, {Schaefer}, {Short}, {Smale}, {Smith},
  {Stella}, {Tagliaferri}, {Takahashi}, {Tashiro}, {Townsley}, {Tueller},
  {Turner}, {Vietri}, {Voges}, {Ward}, {Willingale}, {Zerbi}, \&
  {Zhang}}]{Gehrels04}
{Gehrels}, N., {Chincarini}, G., {Giommi}, P., {et~al.} 2004, \apj, 611, 1005,
  \dodoi{10.1086/422091}

\bibitem[{{Gezari}(2021)}]{Gezari2021TDE}
{Gezari}, S. 2021, \araa, 59, 21, \dodoi{10.1146/annurev-astro-111720-030029}

\bibitem[{{Gezari} {et~al.}(2012){Gezari}, {Chornock}, {Rest}, {Huber},
  {Forster}, {Berger}, {Challis}, {Neill}, {Martin}, {Heckman}, {Lawrence},
  {Norman}, {Narayan}, {Foley}, {Marion}, {Scolnic}, {Chomiuk}, {Soderberg},
  {Smith}, {Kirshner}, {Riess}, {Smartt}, {Stubbs}, {Tonry}, {Wood-Vasey},
  {Burgett}, {Chambers}, {Grav}, {Heasley}, {Kaiser}, {Kudritzki}, {Magnier},
  {Morgan}, \& {Price}}]{Gezari2012Nature}
{Gezari}, S., {Chornock}, R., {Rest}, A., {et~al.} 2012, \nat, 485, 217,
  \dodoi{10.1038/nature10990}

\bibitem[{{Gezari} {et~al.}(2017){Gezari}, {Hung}, {Cenko}, {Blagorodnova},
  {Yan}, {Kulkarni}, {Mooley}, {Kong}, {Cantwell}, {Yu}, {Cao}, {Fremling},
  {Neill}, {Ngeow}, {Nugent}, \& {Wozniak}}]{Gezari2017Rapid}
{Gezari}, S., {Hung}, T., {Cenko}, S.~B., {et~al.} 2017, \apj, 835, 144,
  \dodoi{10.3847/1538-4357/835/2/144}

\bibitem[{{Gomez} {et~al.}(2020){Gomez}, {Nicholl}, {Short}, {Margutti},
  {Alexander}, {Blanchard}, {Berger}, {Eftekhari}, {Schulze}, {Anderson},
  {Arcavi}, {Chornock}, {Cowperthwaite}, {Galbany}, {Herzog}, {Hiramatsu},
  {Hosseinzadeh}, {Laskar}, {M{\"u}ller Bravo}, {Patton}, \&
  {Terreran}}]{Gomez2020AT2018hyz}
{Gomez}, S., {Nicholl}, M., {Short}, P., {et~al.} 2020, \mnras, 497, 1925,
  \dodoi{10.1093/mnras/staa2099}

\bibitem[{{Grayling} {et~al.}(2022){Grayling}, {Toy}, {Wiseman}, {Frohmaier},
  {Galbany}, {Onori}, {Kravtsov}, {Benetti}, {Anderson}, {Chen}, {Gromadzki},
  {Inserra}, {Kankare}, {Bravo}, {Nicholl}, {Yaron}, {Young}, {Zimmerman},
  {Tonry}, {Denneau}, {Heinze}, {Weiland}, {Stalder}, {Rest}, {Smith},
  {Smartt}, {Fulton}, {Gillanders}, {Moore}, \& {Srivastav}}]{BarbieTNS}
{Grayling}, M., {Toy}, M., {Wiseman}, P., {et~al.} 2022, Transient Name Server
  AstroNote, 195, 1

\bibitem[{{Guillochon} {et~al.}(2018){Guillochon}, {Nicholl}, {Villar},
  {Mockler}, {Narayan}, {Mandel}, {Berger}, \& {Williams}}]{Guillochon2018}
{Guillochon}, J., {Nicholl}, M., {Villar}, V.~A., {et~al.} 2018, \apjs, 236, 6,
  \dodoi{10.3847/1538-4365/aab761}

\bibitem[{{Heckman} {et~al.}(2004){Heckman}, {Kauffmann}, {Brinchmann},
  {Charlot}, {Tremonti}, \& {White}}]{heckman04}
{Heckman}, T.~M., {Kauffmann}, G., {Brinchmann}, J., {et~al.} 2004, \apj, 613,
  109, \dodoi{10.1086/422872}

\bibitem[{{Heger} \& {Woosley}(2002)}]{HW02}
{Heger}, A., \& {Woosley}, S.~E. 2002, \apj, 567, 532, \dodoi{10.1086/338487}

\bibitem[{{Hinkle} {et~al.}(2022){Hinkle}, {Holoien}, {Shappee}, {Neustadt},
  {Auchettl}, {Vallely}, {Shahbandeh}, {Kluge}, {Kochanek}, {Stanek}, {Huber},
  {Post}, {Bersier}, {Ashall}, {Tucker}, {Williams}, {de Jaeger}, {Do},
  {Fausnaugh}, {Gruen}, {Hopp}, {Myles}, {Obermeier}, {Payne}, \&
  {Thompson}}]{Hinkle2022ASASSN-20hx}
{Hinkle}, J.~T., {Holoien}, T. W.~S., {Shappee}, B.~J., {et~al.} 2022, \apj,
  930, 12, \dodoi{10.3847/1538-4357/ac5f54}

\bibitem[{{Hodgkin} {et~al.}(2019){Hodgkin}, {Breedt}, {Delgado}, {Harrison},
  {Leeuwen}, {Rixon}, {Wevers}, {Yoldas}, {Ihanec}, {Kruszy{\'n}ska},
  {Rybicki}, {Wyrzykowski}, {Kostrzewa-Rutkowska}, {Eappachen}, \&
  {Marton}}]{Hodgkin2019}
{Hodgkin}, S.~T., {Breedt}, E., {Delgado}, A., {et~al.} 2019, Transient Name
  Server Discovery Report, 2019-2344, 1

\bibitem[{{Holoien} {et~al.}(2014){Holoien}, {Prieto}, {Bersier}, {Kochanek},
  {Stanek}, {Shappee}, {Grupe}, {Basu}, {Beacom}, {Brimacombe}, {Brown},
  {Davis}, {Jencson}, {Pojmanski}, \& {Szczygie{\l}}}]{Holoien2014ASASSN14ae}
{Holoien}, T.~W.~S., {Prieto}, J.~L., {Bersier}, D., {et~al.} 2014, \mnras,
  445, 3263, \dodoi{10.1093/mnras/stu1922}

\bibitem[{{Holoien} {et~al.}(2022){Holoien}, {Neustadt}, {Vallely}, {Auchettl},
  {Hinkle}, {Romero-Ca{\~n}izales}, {Shappee}, {Kochanek}, {Stanek}, {Chen},
  {Dong}, {Prieto}, {Thompson}, {Brink}, {Filippenko}, {Zheng}, {Bersier},
  {Bose}, {Burgasser}, {Channa}, {de Jaeger}, {Hestenes}, {Im}, {Jeffers},
  {Jun}, {Lansbury}, {Post}, {Ross}, {Stern}, {Tang}, {Tucker}, {Valenti},
  {Yunus}, \& {Zhang}}]{Holoien2022asassn-17jz}
{Holoien}, T. W.~S., {Neustadt}, J. M.~M., {Vallely}, P.~J., {et~al.} 2022,
  \apj, 933, 196, \dodoi{10.3847/1538-4357/ac74b9}

\bibitem[{{Huang} \& {Lu}(2022)}]{Huang2023}
{Huang}, H.-T., \& {Lu}, W. 2022, arXiv e-prints, arXiv:2301.00259,
  \dodoi{10.48550/arXiv.2301.00259}

\bibitem[{{Hubeny} \& {Hubeny}(1998)}]{H1988}
{Hubeny}, I., \& {Hubeny}, V. 1998, in American Institute of Physics Conference
  Series, Vol. 431, Accretion processes in Astrophysical Systems: Some like it
  hot! - eigth AstroPhysics Conference, ed. S.~S. {Holt} \& T.~R. {Kallman},
  171--174, \dodoi{10.1063/1.55892}

\bibitem[{{Hung} {et~al.}(2021){Hung}, {Foley}, {Veilleux}, {Cenko}, {Dai},
  {Auchettl}, {Brink}, {Dimitriadis}, {Filippenko}, {Gezari}, {Holoien},
  {Kilpatrick}, {Mockler}, {Piro}, {Ramirez-Ruiz}, {Rojas-Bravo}, {Siebert},
  {van Velzen}, \& {Zheng}}]{Hung2021}
{Hung}, T., {Foley}, R.~J., {Veilleux}, S., {et~al.} 2021, \apj, 917, 9,
  \dodoi{10.3847/1538-4357/abf4c3}

\bibitem[{{IRSA}(2022)}]{IRSA539}
{IRSA}. 2022, Zwicky Transient Facility Image Service,  IPAC,
  \dodoi{10.26131/IRSA539}

\bibitem[{{Kalberla} {et~al.}(2005){Kalberla}, {Burton}, {Hartmann}, {Arnal},
  {Bajaja}, {Morras}, \& {P{\"o}ppel}}]{Kalberla05}
{Kalberla}, P.~M.~W., {Burton}, W.~B., {Hartmann}, D., {et~al.} 2005, \aap,
  440, 775, \dodoi{10.1051/0004-6361:20041864}

\bibitem[{{Kasen} \& {Bildsten}(2010)}]{Kasen10}
{Kasen}, D., \& {Bildsten}, L. 2010, \apj, 717, 245,
  \dodoi{10.1088/0004-637X/717/1/245}

\bibitem[{{Kasen} {et~al.}(2011){Kasen}, {Woosley}, \& {Heger}}]{Kasen11}
{Kasen}, D., {Woosley}, S.~E., \& {Heger}, A. 2011, \apj, 734, 102,
  \dodoi{10.1088/0004-637X/734/2/102}

\bibitem[{{Kiewe} {et~al.}(2012){Kiewe}, {Gal-Yam}, {Arcavi}, {Leonard},
  {Emilio Enriquez}, {Cenko}, {Fox}, {Moon}, {Sand}, {Soderberg}, \&
  {CCCP}}]{Kiewe2012SN2005cl}
{Kiewe}, M., {Gal-Yam}, A., {Arcavi}, I., {et~al.} 2012, \apj, 744, 10,
  \dodoi{10.1088/0004-637X/744/1/10}

\bibitem[{{Kochanek}(2016)}]{kochanek16}
{Kochanek}, C.~S. 2016, \mnras, 458, 127, \dodoi{10.1093/mnras/stw267}

\bibitem[{{Kormendy} \& {Ho}(2013)}]{KH13}
{Kormendy}, J., \& {Ho}, L.~C. 2013, \araa, 51, 511,
  \dodoi{10.1146/annurev-astro-082708-101811}

\bibitem[{{Lacy} {et~al.}(2020){Lacy}, {Baum}, {Chandler}, {Chatterjee},
  {Clarke}, {Deustua}, {English}, {Farnes}, {Gaensler}, {Gugliucci},
  {Hallinan}, {Kent}, {Kimball}, {Law}, {Lazio}, {Marvil}, {Mao}, {Medlin},
  {Mooley}, {Murphy}, {Myers}, {Osten}, {Richards}, {Rosolowsky}, {Rudnick},
  {Schinzel}, {Sivakoff}, {Sjouwerman}, {Taylor}, {White}, {Wrobel},
  {Andernach}, {Beasley}, {Berger}, {Bhatnager}, {Birkinshaw}, {Bower},
  {Brandt}, {Brown}, {Burke-Spolaor}, {Butler}, {Comerford}, {Demorest}, {Fu},
  {Giacintucci}, {Golap}, {G{\"u}th}, {Hales}, {Hiriart}, {Hodge}, {Horesh},
  {Ivezi{\'c}}, {Jarvis}, {Kamble}, {Kassim}, {Liu}, {Loinard}, {Lyons},
  {Masters}, {Mezcua}, {Moellenbrock}, {Mroczkowski}, {Nyland}, {O'Dea},
  {O'Sullivan}, {Peters}, {Radford}, {Rao}, {Robnett}, {Salcido}, {Shen},
  {Sobotka}, {Witz}, {Vaccari}, {van Weeren}, {Vargas}, {Williams}, \&
  {Yoon}}]{Lacy20}
{Lacy}, M., {Baum}, S.~A., {Chandler}, C.~J., {et~al.} 2020, \pasp, 132,
  035001, \dodoi{10.1088/1538-3873/ab63eb}

\bibitem[{{Leloudas} {et~al.}(2016){Leloudas}, {Fraser}, {Stone}, {van Velzen},
  {Jonker}, {Arcavi}, {Fremling}, {Maund}, {Smartt}, {Kr{\`\i}hler},
  {Miller-Jones}, {Vreeswijk}, {Gal-Yam}, {Mazzali}, {De Cia}, {Howell},
  {Inserra}, {Patat}, {de Ugarte Postigo}, {Yaron}, {Ashall}, {Bar},
  {Campbell}, {Chen}, {Childress}, {Elias-Rosa}, {Harmanen}, {Hosseinzadeh},
  {Johansson}, {Kangas}, {Kankare}, {Kim}, {Kuncarayakti}, {Lyman}, {Magee},
  {Maguire}, {Malesani}, {Mattila}, {McCully}, {Nicholl}, {Prentice},
  {Romero-Ca{\~n}izales}, {Schulze}, {Smith}, {Sollerman}, {Sullivan},
  {Tucker}, {Valenti}, {Wheeler}, \& {Young}}]{Leloudas2016}
{Leloudas}, G., {Fraser}, M., {Stone}, N.~C., {et~al.} 2016, Nature Astronomy,
  1, 0002, \dodoi{10.1038/s41550-016-0002}

\bibitem[{{Leloudas} {et~al.}(2019){Leloudas}, {Dai}, {Arcavi}, {Vreeswijk},
  {Mockler}, {Roy}, {Malesani}, {Schulze}, {Wevers}, {Fraser}, {Ramirez-Ruiz},
  {Auchettl}, {Burke}, {Cannizzaro}, {Charalampopoulos}, {Chen}, {Cikota},
  {Della Valle}, {Galbany}, {Gromadzki}, {Heintz}, {Hiramatsu}, {Jonker},
  {Kostrzewa-Rutkowska}, {Maguire}, {Mandel}, {Nicholl}, {Onori}, {Roth},
  {Smartt}, {Wyrzykowski}, \& {Young}}]{Leloudas2019}
{Leloudas}, G., {Dai}, L., {Arcavi}, I., {et~al.} 2019, \apj, 887, 218,
  \dodoi{10.3847/1538-4357/ab5792}

\bibitem[{{Li} {et~al.}(2019){Li}, {Wang}, {Hu}, {Yang}, {Zhang}, {Mo}, {Chen},
  {Zhang}, {Benetti}, {Cappellaro}, {Elias-Rosa}, {Isern}, {Morales-Garoffolo},
  {Huang}, {Ochner}, {Pastorello}, {Reguitti}, {Tartaglia}, {Terreran},
  {Tomasella}, \& {Wang}}]{Li2019SN2014J}
{Li}, W., {Wang}, X., {Hu}, M., {et~al.} 2019, \apj, 882, 30,
  \dodoi{10.3847/1538-4357/ab2b49}

\bibitem[{{Mancone} \& {Gonzalez}(2012)}]{Mancone2012Ez}
{Mancone}, C.~L., \& {Gonzalez}, A.~H. 2012, \pasp, 124, 606,
  \dodoi{10.1086/666502}

\bibitem[{{Margutti} {et~al.}(2013){Margutti}, {Zaninoni}, {Bernardini},
  {Chincarini}, {Pasotti}, {Guidorzi}, {Angelini}, {Burrows}, {Capalbi},
  {Evans}, {Gehrels}, {Kennea}, {Mangano}, {Moretti}, {Nousek}, {Osborne},
  {Page}, {Perri}, {Racusin}, {Romano}, {Sbarufatti}, {Stafford}, \&
  {Stamatikos}}]{Margutti13}
{Margutti}, R., {Zaninoni}, E., {Bernardini}, M.~G., {et~al.} 2013, \mnras,
  428, 729, \dodoi{10.1093/mnras/sts066}

\bibitem[{{Margutti} {et~al.}(2017){Margutti}, {Metzger}, {Chornock},
  {Milisavljevic}, {Berger}, {Blanchard}, {Guidorzi}, {Migliori}, {Kamble},
  {Lunnan}, {Nicholl}, {Coppejans}, {Dall'Osso}, {Drout}, {Perna}, \&
  {Sbarufatti}}]{Margutti2017ASASSN15lh}
{Margutti}, R., {Metzger}, B.~D., {Chornock}, R., {et~al.} 2017, \apj, 836, 25,
  \dodoi{10.3847/1538-4357/836/1/25}

\bibitem[{{Mart{\'\i}n-Navarro} {et~al.}(2018){Mart{\'\i}n-Navarro}, {Brodie},
  {Romanowsky}, {Ruiz-Lara}, \& {van de Ven}}]{Martin2018Nature}
{Mart{\'\i}n-Navarro}, I., {Brodie}, J.~P., {Romanowsky}, A.~J., {Ruiz-Lara},
  T., \& {van de Ven}, G. 2018, \nat, 553, 307, \dodoi{10.1038/nature24999}

\bibitem[{{Masci} {et~al.}(2019){Masci}, {Laher}, {Rusholme}, {Shupe}, {Groom},
  {Surace}, {Jackson}, {Monkewitz}, {Beck}, {Flynn}, {Terek}, {Landry},
  {Hacopians}, {Desai}, {Howell}, {Brooke}, {Imel}, {Wachter}, {Ye}, {Lin},
  {Cenko}, {Cunningham}, {Rebbapragada}, {Bue}, {Miller}, {Mahabal}, {Bellm},
  {Patterson}, {Juri{\'c}}, {Golkhou}, {Ofek}, {Walters}, {Graham}, {Kasliwal},
  {Dekany}, {Kupfer}, {Burdge}, {Cannella}, {Barlow}, {Van Sistine}, {Giomi},
  {Fremling}, {Blagorodnova}, {Levitan}, {Riddle}, {Smith}, {Helou}, {Prince},
  \& {Kulkarni}}]{Masci2019}
{Masci}, F.~J., {Laher}, R.~R., {Rusholme}, B., {et~al.} 2019, \pasp, 131,
  018003, \dodoi{10.1088/1538-3873/aae8ac}

\bibitem[{{Matheson} {et~al.}(2021){Matheson}, {Stubens}, {Wolf}, {Lee},
  {Narayan}, {Saha}, {Scott}, {Soraisam}, {Bolton}, {Hauger}, {Silva},
  {Kececioglu}, {Scheidegger}, {Snodgrass}, {Aleo}, {Evans-Jacquez}, {Singh},
  {Wang}, {Yang}, \& {Zhao}}]{Matheson21}
{Matheson}, T., {Stubens}, C., {Wolf}, N., {et~al.} 2021, \aj, 161, 107,
  \dodoi{10.3847/1538-3881/abd703}

\bibitem[{{McConnell} \& {Ma}(2013)}]{MP13}
{McConnell}, N.~J., \& {Ma}, C.-P. 2013, \apj, 764, 184,
  \dodoi{10.1088/0004-637X/764/2/184}

\bibitem[{{McKinney} \& {Blandford}(2009)}]{M2009Jets}
{McKinney}, J.~C., \& {Blandford}, R.~D. 2009, \mnras, 394, L126,
  \dodoi{10.1111/j.1745-3933.2009.00625.x}

\bibitem[{Miller \& Stone(1993)}]{miller93}
Miller, J., \& Stone, R. 1993, Lick Obs., Santa Cruz, CA, 66

\bibitem[{{Mockler} {et~al.}(2019){Mockler}, {Guillochon}, \&
  {Ramirez-Ruiz}}]{Mockler2019TDE}
{Mockler}, B., {Guillochon}, J., \& {Ramirez-Ruiz}, E. 2019, \apj, 872, 151,
  \dodoi{10.3847/1538-4357/ab010f}

\bibitem[{{Mushotzky}(1988)}]{M1988}
{Mushotzky}, R.~F. 1988, in Active Galactic Nuclei, ed. H.~R. {Miller} \& P.~J.
  {Wiita}, Vol. 307, 239, \dodoi{10.1007/3-540-19492-$4_204$}

\bibitem[{{Neustadt} {et~al.}(2020){Neustadt}, {Holoien}, {Kochanek},
  {Auchettl}, {Brown}, {Shappee}, {Pogge}, {Dong}, {Stanek}, {Tucker}, {Bose},
  {Chen}, {Ricci}, {Vallely}, {Prieto}, {Thompson}, {Coulter}, {Drout},
  {Foley}, {Kilpatrick}, {Piro}, {Rojas-Bravo}, {Buckley}, {Gromadzki},
  {Dimitriadis}, {Siebert}, {Do}, {Huber}, \& {Payne}}]{Neustadt2020ASASSN18jd}
{Neustadt}, J.~M.~M., {Holoien}, T.~W.~S., {Kochanek}, C.~S., {et~al.} 2020,
  \mnras, 494, 2538, \dodoi{10.1093/mnras/staa859}

\bibitem[{{Nicholl} {et~al.}(2022){Nicholl}, {Lanning}, {Ramsden}, {Mockler},
  {Lawrence}, {Short}, \& {Ridley}}]{Nicoll2022}
{Nicholl}, M., {Lanning}, D., {Ramsden}, P., {et~al.} 2022, \mnras, 515, 5604,
  \dodoi{10.1093/mnras/stac2206}

\bibitem[{{Nicholl} {et~al.}(2020){Nicholl}, {Wevers}, {Oates}, {Alexander},
  {Leloudas}, {Onori}, {Jerkstrand}, {Gomez}, {Campana}, {Arcavi},
  {Charalampopoulos}, {Gromadzki}, {Ihanec}, {Jonker}, {Lawrence}, {Mandel},
  {Schulze}, {Short}, {Burke}, {McCully}, {Hiramatsu}, {Howell}, {Pellegrino},
  {Abbot}, {Anderson}, {Berger}, {Blanchard}, {Cannizzaro}, {Chen},
  {Dennefeld}, {Galbany}, {Gonz{\'a}lez-Gait{\'a}n}, {Hosseinzadeh}, {Inserra},
  {Irani}, {Kuin}, {M{\"u}ller-Bravo}, {Pineda}, {Ross}, {Roy}, {Smartt},
  {Smith}, {Tucker}, {Wyrzykowski}, \& {Young}}]{Nicoll2020}
{Nicholl}, M., {Wevers}, T., {Oates}, S.~R., {et~al.} 2020, \mnras, 499, 482,
  \dodoi{10.1093/mnras/staa2824}

\bibitem[{Oke \& Gunn(1982)}]{Oke_1982}
Oke, J.~B., \& Gunn, J.~E. 1982, Publications of the Astronomical Society of
  the Pacific, 94, 586, \dodoi{10.1086/131027}

\bibitem[{{Oke} {et~al.}(1995){Oke}, {Cohen}, {Carr}, {Cromer}, {Dingizian},
  {Harris}, {Labrecque}, {Lucinio}, {Schaal}, {Epps}, \& {Miller}}]{Oke95}
{Oke}, J.~B., {Cohen}, J.~G., {Carr}, M., {et~al.} 1995, \pasp, 107, 375,
  \dodoi{10.1086/133562}

\bibitem[{{Oknyanskij}(1978)}]{Oknyanskij1978}
{Oknyanskij}, V.~L. 1978, Peremennye Zvezdy, 21, 71

\bibitem[{{Osterbrock} \& {Ferland}(2006)}]{agn2}
{Osterbrock}, D.~E., \& {Ferland}, G.~J. 2006, {Astrophysics of gaseous nebulae
  and active galactic nuclei}

\bibitem[{{Pasham} {et~al.}(2015){Pasham}, {Cenko}, {Levan}, {Bower}, {Horesh},
  {Brown}, {Dolan}, {Wiersema}, {Filippenko}, {Fruchter}, {Greiner}, {O'Brien},
  {Page}, {Rau}, \& {Tanvir}}]{Pasham2015}
{Pasham}, D.~R., {Cenko}, S.~B., {Levan}, A.~J., {et~al.} 2015, \apj, 805, 68,
  \dodoi{10.1088/0004-637X/805/1/68}

\bibitem[{{Peterson} {et~al.}(2004){Peterson}, {Ferrarese}, {Gilbert}, {Kaspi},
  {Malkan}, {Maoz}, {Merritt}, {Netzer}, {Onken}, {Pogge}, {Vestergaard}, \&
  {Wandel}}]{Peterson2004}
{Peterson}, B.~M., {Ferrarese}, L., {Gilbert}, K.~M., {et~al.} 2004, \apj, 613,
  682, \dodoi{10.1086/423269}

\bibitem[{{Peterson} {et~al.}(2013){Peterson}, {Denney}, {De Rosa}, {Grier},
  {Pogge}, {Bentz}, {Kochanek}, {Vestergaard}, {Kilerci-Eser}, {Dalla
  Bont{\`a}}, \& {Ciroi}}]{peterson2013}
{Peterson}, B.~M., {Denney}, K.~D., {De Rosa}, G., {et~al.} 2013, \apj, 779,
  109, \dodoi{10.1088/0004-637X/779/2/109}

\bibitem[{{Polzin} {et~al.}(2022){Polzin}, {Margutti}, {Coppejans}, {Auchettl},
  {Page}, {Vasilopoulos}, {Bright}, {Esposito}, {Williams}, {Mukai}, \&
  {Berger}}]{Polzin22}
{Polzin}, A., {Margutti}, R., {Coppejans}, D., {et~al.} 2022, arXiv e-prints,
  arXiv:2211.01232, \dodoi{10.48550/arXiv.2211.01232}

\bibitem[{{Rees}(1988)}]{Rees1988tde}
{Rees}, M.~J. 1988, \nat, 333, 523, \dodoi{10.1038/333523a0}

\bibitem[{{Ricci} {et~al.}(2020){Ricci}, {Kara}, {Loewenstein}, {Trakhtenbrot},
  {Arcavi}, {Remillard}, {Fabian}, {Gendreau}, {Arzoumanian}, {Li}, {Ho},
  {MacLeod}, {Cackett}, {Altamirano}, {Gandhi}, {Kosec}, {Pasham}, {Steiner},
  \& {Chan}}]{Ricci2020}
{Ricci}, C., {Kara}, E., {Loewenstein}, M., {et~al.} 2020, \apjl, 898, L1,
  \dodoi{10.3847/2041-8213/ab91a1}

\bibitem[{{Roming} {et~al.}(2005){Roming}, {Kennedy}, {Mason}, {Nousek}, {Ahr},
  {Bingham}, {Broos}, {Carter}, {Hancock}, {Huckle}, {Hunsberger}, {Kawakami},
  {Killough}, {Koch}, {McLelland}, {Smith}, {Smith}, {Soto}, {Boyd},
  {Breeveld}, {Holland}, {Ivanushkina}, {Pryzby}, {Still}, \&
  {Stock}}]{Roming05}
{Roming}, P. W.~A., {Kennedy}, T.~E., {Mason}, K.~O., {et~al.} 2005, \ssr, 120,
  95, \dodoi{10.1007/s11214-005-5095-4}

\bibitem[{{Schlafly} \& {Finkbeiner}(2011)}]{Schlafly2011}
{Schlafly}, E.~F., \& {Finkbeiner}, D.~P. 2011, \apj, 737, 103,
  \dodoi{10.1088/0004-637X/737/2/103}

\bibitem[{{Schmidt} {et~al.}(2018){Schmidt}, {Oio}, {Ferreiro}, {Vega}, \&
  {Weidmann}}]{Schmidt2018}
{Schmidt}, E.~O., {Oio}, G.~A., {Ferreiro}, D., {Vega}, L., \& {Weidmann}, W.
  2018, \aap, 615, A13, \dodoi{10.1051/0004-6361/201731557}

\bibitem[{{Science Software Branch at STScI}(2012)}]{Pyraf2012}
{Science Software Branch at STScI}. 2012, {PyRAF: Python alternative for IRAF},
  Astrophysics Source Code Library, record ascl:1207.011.
\newblock \doeprint{1207.011}

\bibitem[{{Shappee} {et~al.}(2014){Shappee}, {Prieto}, {Grupe}, {Kochanek},
  {Stanek}, {De Rosa}, {Mathur}, {Zu}, {Peterson}, {Pogge}, {Komossa}, {Im},
  {Jencson}, {Holoien}, {Basu}, {Beacom}, {Szczygie{\l}}, {Brimacombe},
  {Adams}, {Campillay}, {Choi}, {Contreras}, {Dietrich}, {Dubberley},
  {Elphick}, {Foale}, {Giustini}, {Gonzalez}, {Hawkins}, {Howell}, {Hsiao},
  {Koss}, {Leighly}, {Morrell}, {Mudd}, {Mullins}, {Nugent}, {Parrent},
  {Phillips}, {Pojmanski}, {Rosing}, {Ross}, {Sand}, {Terndrup}, {Valenti},
  {Walker}, \& {Yoon}}]{Shappee2014Outburst}
{Shappee}, B.~J., {Prieto}, J.~L., {Grupe}, D., {et~al.} 2014, \apj, 788, 48,
  \dodoi{10.1088/0004-637X/788/1/48}

\bibitem[{Skilling(2006)}]{Skilling2006}
Skilling, J. 2006, Bayesian Analysis, 1, 833 , \dodoi{10.1214/06-BA127}

\bibitem[{{Smartt} {et~al.}(2015){Smartt}, {Valenti}, {Fraser}, {Inserra},
  {Young}, {Sullivan}, {Pastorello}, {Benetti}, {Gal-Yam}, {Knapic},
  {Molinaro}, {Smareglia}, {Smith}, {Taubenberger}, {Yaron}, {Anderson},
  {Ashall}, {Balland}, {Baltay}, {Barbarino}, {Bauer}, {Baumont}, {Bersier},
  {Blagorodnova}, {Bongard}, {Botticella}, {Bufano}, {Bulla}, {Cappellaro},
  {Campbell}, {Cellier-Holzem}, {Chen}, {Childress}, {Clocchiatti},
  {Contreras}, {Dall'Ora}, {Danziger}, {de Jaeger}, {De Cia}, {Della Valle},
  {Dennefeld}, {Elias-Rosa}, {Elman}, {Feindt}, {Fleury}, {Gall},
  {Gonzalez-Gaitan}, {Galbany}, {Morales Garoffolo}, {Greggio}, {Guillou},
  {Hachinger}, {Hadjiyska}, {Hage}, {Hillebrandt}, {Hodgkin}, {Hsiao}, {James},
  {Jerkstrand}, {Kangas}, {Kankare}, {Kotak}, {Kromer}, {Kuncarayakti},
  {Leloudas}, {Lundqvist}, {Lyman}, {Hook}, {Maguire}, {Manulis}, {Margheim},
  {Mattila}, {Maund}, {Mazzali}, {McCrum}, {McKinnon}, {Moreno-Raya},
  {Nicholl}, {Nugent}, {Pain}, {Pignata}, {Phillips}, {Polshaw}, {Pumo},
  {Rabinowitz}, {Reilly}, {Romero-Ca{\~n}izales}, {Scalzo}, {Schmidt},
  {Schulze}, {Sim}, {Sollerman}, {Taddia}, {Tartaglia}, {Terreran},
  {Tomasella}, {Turatto}, {Walker}, {Walton}, {Wyrzykowski}, {Yuan}, \&
  {Zampieri}}]{Smartt2015}
{Smartt}, S.~J., {Valenti}, S., {Fraser}, M., {et~al.} 2015, \aap, 579, A40,
  \dodoi{10.1051/0004-6361/201425237}

\bibitem[{{Smith} {et~al.}(2018){Smith}, {Mushotzky}, {Boyd}, {Malkan},
  {Howell}, \& {Gelino}}]{Smith2018}
{Smith}, K.~L., {Mushotzky}, R.~F., {Boyd}, P.~T., {et~al.} 2018, \apj, 857,
  141, \dodoi{10.3847/1538-4357/aab88d}

\bibitem[{{Smith} {et~al.}(2020){Smith}, {Smartt}, {Young}, {Tonry}, {Denneau},
  {Flewelling}, {Heinze}, {Weiland}, {Stalder}, {Rest}, {Stubbs}, {Anderson},
  {Chen}, {Clark}, {Do}, {F{\"o}rster}, {Fulton}, {Gillanders}, {McBrien},
  {O'Neill}, {Srivastav}, \& {Wright}}]{Smith2020}
{Smith}, K.~W., {Smartt}, S.~J., {Young}, D.~R., {et~al.} 2020, \pasp, 132,
  085002, \dodoi{10.1088/1538-3873/ab936e}

\bibitem[{{Smith} {et~al.}(2007){Smith}, {Li}, {Foley}, {Wheeler}, {Pooley},
  {Chornock}, {Filippenko}, {Silverman}, {Quimby}, {Bloom}, \&
  {Hansen}}]{Nathan2007SN2006gy}
{Smith}, N., {Li}, W., {Foley}, R.~J., {et~al.} 2007, \apj, 666, 1116,
  \dodoi{10.1086/519949}

\bibitem[{{Sravan} {et~al.}(2020){Sravan}, {Milisavljevic}, {Reynolds},
  {Lentner}, \& {Linvill}}]{Sravan2020}
{Sravan}, N., {Milisavljevic}, D., {Reynolds}, J.~M., {Lentner}, G., \&
  {Linvill}, M. 2020, \apj, 893, 127, \dodoi{10.3847/1538-4357/ab8128}

\bibitem[{{Stone} \& {Metzger}(2016)}]{Stone2016}
{Stone}, N.~C., \& {Metzger}, B.~D. 2016, \mnras, 455, 859,
  \dodoi{10.1093/mnras/stv2281}

\bibitem[{{Tonry} {et~al.}(2018){Tonry}, {Denneau}, {Heinze}, {Stalder},
  {Smith}, {Smartt}, {Stubbs}, {Weiland}, \& {Rest}}]{Tonry2018}
{Tonry}, J.~L., {Denneau}, L., {Heinze}, A.~N., {et~al.} 2018, \pasp, 130,
  064505, \dodoi{10.1088/1538-3873/aabadf}

\bibitem[{{Trakhtenbrot} {et~al.}(2019){Trakhtenbrot}, {Arcavi}, {MacLeod},
  {Ricci}, {Kara}, {Graham}, {Stern}, {Harrison}, {Burke}, {Hiramatsu},
  {Hosseinzadeh}, {Howell}, {Smartt}, {Rest}, {Prieto}, {Shappee}, {Holoien},
  {Bersier}, {Filippenko}, {Brink}, {Zheng}, {Li}, {Remillard}, \&
  {Loewenstein}}]{Trakhtenbrot2019}
{Trakhtenbrot}, B., {Arcavi}, I., {MacLeod}, C.~L., {et~al.} 2019, \apj, 883,
  94, \dodoi{10.3847/1538-4357/ab39e4}

\bibitem[{{van Leeuwen} {et~al.}(2018){van Leeuwen}, {de Bruijne}, {Arenou},
  {Bakker}, {Blomme}, {Busso}, {Cacciari}, {Casta{\~n}eda}, {Cellino},
  {Clotet}, {Comoretto}, {Eyer}, {Gonz{\'a}lez-N{\'u}{\~n}ez}, {Guy}, {Hambly},
  {Hobbs}, {van Leeuwen}, {Luri}, {Manteiga}, {Pourbaix}, {Roegiers},
  {Salgado}, {Sartoretti}, {Tanga}, {Ulla}, {Utrilla Molina}, {Abreu},
  {Altmann}, {Andrae}, {Antoja}, {Audard}, {Babusiaux}, {Bailer-Jones},
  {Barache}, {Bastian}, {Beck}, {Berthier}, {Bianchi}, {Biermann}, {Bombrun},
  {Bossini}, {Breddels}, {Brown}, {Busonero}, {Butkevich}, {Cantat-Gaudin},
  {Carrasco}, {Cheek}, {Clementini}, {Creevey}, {Crowley}, {David}, {Davidson},
  {De Angeli}, {De Ridder}, {Delb{\`o}}, {Dell'Oro}, {Diakit{\'e}},
  {Distefano}, {Drimmel}, {Dur{\'a}n}, {Evans}, {Fabricius}, {Fabrizio},
  {Fern{\'a}ndez-Hern{\'a}ndez}, {Findeisen}, {Fleitas}, {Fouesneau},
  {Galluccio}, {Gracia-Abril}, {Guerra}, {Guti{\'e}rrez-S{\'a}nchez}, {Helmi},
  {Hernandez}, {Holl}, {Hutton}, {Jean-Antoine-Piccolo}, {Jevardat de
  Fombelle}, {Joliet}, {Jordi}, {Juh{\'a}sz}, {Klioner}, {L{\"o}ffler},
  {Lammers}, {Lanzafame}, {Lebzelter}, {Leclerc}, {Lecoeur-Ta{\"\i}bi},
  {Lindegren}, {Marinoni}, {Marrese}, {Mary}, {Massari}, {Messineo},
  {Michalik}, {Mignard}, {Molinaro}, {Moln{\'a}r}, {Montegriffo}, {Mora},
  {Mowlavi}, {Muinonen}, {Muraveva}, {Nienartowicz}, {Ordenovic}, {Pancino},
  {Panem}, {Pauwels}, {Petit}, {Plachy}, {Portell}, {Racero}, {Regibo},
  {Reyl{\'e}}, {Rimoldini}, {Ripepi}, {Riva}, {Robichon}, {Robin}, {Roelens},
  {Romero-G{\'o}mez}, {Sarro}, {Seabroke}, {Segovia}, {Siddiqui}, {Smart},
  {Smith}, {Sordo}, {Soria}, {Spoto}, {Stephenson}, {Turon}, {Vallenari},
  {Veljanoski}, \& {Voutsinas}}]{GaiaDR22018}
{van Leeuwen}, F., {de Bruijne}, J.~H.~J., {Arenou}, F., {et~al.} 2018, {Gaia
  DR2 documentation}, Gaia DR2 documentation, European Space Agency; Gaia Data
  Processing and Analysis Consortium.

\bibitem[{{van Velzen} {et~al.}(2021){van Velzen}, {Gezari}, {Hammerstein},
  {Roth}, {Frederick}, {Ward}, {Hung}, {Cenko}, {Stein}, {Perley}, {Taggart},
  {Foley}, {Sollerman}, {Blagorodnova}, {Andreoni}, {Bellm}, {Brinnel}, {De},
  {Dekany}, {Feeney}, {Fremling}, {Giomi}, {Golkhou}, {Graham}, {Ho},
  {Kasliwal}, {Kilpatrick}, {Kulkarni}, {Kupfer}, {Laher}, {Mahabal}, {Masci},
  {Miller}, {Nordin}, {Riddle}, {Rusholme}, {van Santen}, {Sharma}, {Shupe}, \&
  {Soumagnac}}]{Velzen2021}
{van Velzen}, S., {Gezari}, S., {Hammerstein}, E., {et~al.} 2021, \apj, 908, 4,
  \dodoi{10.3847/1538-4357/abc258}

\bibitem[{{Vanden Berk} {et~al.}(2001){Vanden Berk}, {Richards}, {Bauer},
  {Strauss}, {Schneider}, {Heckman}, {York}, {Hall}, {Fan}, {Knapp},
  {Anderson}, {Annis}, {Bahcall}, {Bernardi}, {Briggs}, {Brinkmann}, {Brunner},
  {Burles}, {Carey}, {Castander}, {Connolly}, {Crocker}, {Csabai}, {Doi},
  {Finkbeiner}, {Friedman}, {Frieman}, {Fukugita}, {Gunn}, {Hennessy},
  {Ivezi{\'c}}, {Kent}, {Kunszt}, {Lamb}, {Leger}, {Long}, {Loveday}, {Lupton},
  {Meiksin}, {Merelli}, {Munn}, {Newberg}, {Newcomb}, {Nichol}, {Owen}, {Pier},
  {Pope}, {Rockosi}, {Schlegel}, {Siegmund}, {Smee}, {Snir}, {Stoughton},
  {Stubbs}, {SubbaRao}, {Szalay}, {Szokoly}, {Tremonti}, {Uomoto}, {Waddell},
  {Yanny}, \& {Zheng}}]{Vanden2001}
{Vanden Berk}, D.~E., {Richards}, G.~T., {Bauer}, A., {et~al.} 2001, \aj, 122,
  549, \dodoi{10.1086/321167}

\bibitem[{{Vreeswijk} {et~al.}(2017){Vreeswijk}, {Leloudas}, {Gal-Yam}, {De
  Cia}, {Perley}, {Quimby}, {Waldman}, {Sullivan}, {Yan}, {Ofek}, {Fremling},
  {Taddia}, {Sollerman}, {Valenti}, {Arcavi}, {Howell}, {Filippenko}, {Cenko},
  {Yaron}, {Kasliwal}, {Cao}, {Ben-Ami}, {Horesh}, {Rubin}, {Lunnan}, {Nugent},
  {Laher}, {Rebbapragada}, {Wo{\'z}niak}, \&
  {Kulkarni}}]{Vreeswijk2017PTF12dam}
{Vreeswijk}, P.~M., {Leloudas}, G., {Gal-Yam}, A., {et~al.} 2017, \apj, 835,
  58, \dodoi{10.3847/1538-4357/835/1/58}

\bibitem[{{Wevers} {et~al.}(2017){Wevers}, {van Velzen}, {Jonker}, {Stone},
  {Hung}, {Onori}, {Gezari}, \& {Blagorodnova}}]{Weavers2017}
{Wevers}, T., {van Velzen}, S., {Jonker}, P.~G., {et~al.} 2017, \mnras, 471,
  1694, \dodoi{10.1093/mnras/stx1703}

\bibitem[{{Wiseman} {et~al.}(2023){Wiseman}, {Wang}, {H{\"o}nig},
  {Castro-Segura}, {Clark}, {Frohmaier}, {Fulton}, {Leloudas}, {Middleton},
  {M{\"u}ller-Bravo}, {Mummery}, {Pursiainen}, {Smartt}, {Smith}, {Sullivan},
  {Anderson}, {Acosta Pulido}, {Charalampopoulos}, {Banerji}, {Dennefeld},
  {Galbany}, {Gromadzki}, {Guti{\'e}rrez}, {Ihanec}, {Kankare}, {Lawrence},
  {Mockler}, {Moore}, {Nicholl}, {Onori}, {Petrushevska}, {Ragosta}, {Rest},
  {Smith}, {Wevers}, {Carini}, {Chen}, {Chambers}, {Gao}, {Huber}, {Inserra},
  {Magnier}, {Makrygianni}, {Toy}, {Vincentelli}, \& {Young}}]{Wiseman2023}
{Wiseman}, P., {Wang}, Y., {H{\"o}nig}, S., {et~al.} 2023, arXiv e-prints,
  arXiv:2303.04412, \dodoi{10.48550/arXiv.2303.04412}

\bibitem[{{Woosley}(2017)}]{Woosley2017PPISNe}
{Woosley}, S.~E. 2017, \apj, 836, 244, \dodoi{10.3847/1538-4357/836/2/244}

\bibitem[{{Yan} {et~al.}(2019){Yan}, {Wang}, {Jiang}, {Stern}, {Dou},
  {Fremling}, {Graham}, {Drake}, {Yang}, {Burdge}, \&
  {Kasliwal}}]{Yan2019Rapid}
{Yan}, L., {Wang}, T., {Jiang}, N., {et~al.} 2019, \apj, 874, 44,
  \dodoi{10.3847/1538-4357/ab074b}

\bibitem[{{Yu} {et~al.}(2022){Yu}, {Kochanek}, {Mathur}, {Auchettl}, {Grupe},
  \& {Holoien}}]{Yu2022}
{Yu}, Z., {Kochanek}, C.~S., {Mathur}, S., {et~al.} 2022, \mnras, 515, 5198,
  \dodoi{10.1093/mnras/stac2073}

\bibitem[{{Yusef-Zadeh} {et~al.}(2013){Yusef-Zadeh}, {Royster}, {Wardle},
  {Arendt}, {Bushouse}, {Lis}, {Pound}, {Roberts}, {Whitney}, \&
  {Wootten}}]{Zadeh2013}
{Yusef-Zadeh}, F., {Royster}, M., {Wardle}, M., {et~al.} 2013, \apjl, 767, L32,
  \dodoi{10.1088/2041-8205/767/2/L32}

\bibitem[{{Zauderer} {et~al.}(2011){Zauderer}, {Berger}, {Soderberg}, {Loeb},
  {Narayan}, {Frail}, {Petitpas}, {Brunthaler}, {Chornock}, {Carpenter},
  {Pooley}, {Mooley}, {Kulkarni}, {Margutti}, {Fox}, {Nakar}, {Patel},
  {Volgenau}, {Culverhouse}, {Bietenholz}, {Rupen}, {Max-Moerbeck}, {Readhead},
  {Richards}, {Shepherd}, {Storm}, \& {Hull}}]{Zauderer2011}
{Zauderer}, B.~A., {Berger}, E., {Soderberg}, A.~M., {et~al.} 2011, \nat, 476,
  425, \dodoi{10.1038/nature10366}

\bibitem[{{Zavala} {et~al.}(2018){Zavala}, {Aretxaga}, {Dunlop},
  {Micha{\l}owski}, {Hughes}, {Bourne}, {Chapin}, {Cowley}, {Farrah}, {Lacey},
  {Targett}, \& {van der Werf}}]{Zavala2018}
{Zavala}, J.~A., {Aretxaga}, I., {Dunlop}, J.~S., {et~al.} 2018, \mnras, 475,
  5585, \dodoi{10.1093/mnras/sty217}

\bibitem[{{Zdziarski} {et~al.}(2020){Zdziarski}, {Szanecki}, {Poutanen},
  {Gierli{\'n}ski}, \& {Biernacki}}]{Z2020}
{Zdziarski}, A.~A., {Szanecki}, M., {Poutanen}, J., {Gierli{\'n}ski}, M., \&
  {Biernacki}, P. 2020, \mnras, 492, 5234, \dodoi{10.1093/mnras/staa159}

\end{thebibliography}
